\documentclass[12pt,preprint]{aastex}

\begin{document} 
\title{Further Definition of the Mass-Metallicity Relation in\\ 
Globular Cluster Systems Around Brightest Cluster Galaxies}

\author{Robert Cockcroft}
\affil{Department of Physics and Astronomy, McMaster University, Hamilton, Ontario, L8S 4M1, Canada}
\email{cockcroft@physics.mcmaster.ca}

\author{William E. Harris}
\affil{Department of Physics and Astronomy, McMaster University, Hamilton, Ontario, L8S 4M1, Canada}
\email{harris@physics.mcmaster.ca}

\author{Elizabeth M. H. Wehner}
\affil{Utrecht University, PO Box 80125, 3508 TC Utrecht, The Netherlands}
\email{e.m.wehner@uu.nl}

\author{Bradley C. Whitmore}
\affil{Space Telescope Science Institute, 3700 San Martin Drive,
  Baltimore MD 21218}
\email{whitmore@stsci.edu}

\and

\author{Barry Rothberg}
\affil{Naval Research Laboratory, Code 7211, 4555 Overlook Ave SW, Washington
D.C. 20375}
\email{rothberg@nrl.navy.mil}


\begin{abstract}
We combine the globular cluster data for 
fifteen Brightest Cluster Galaxies and use this material to trace the
mass-metallicity relations (MMR) in their globular
cluster systems (GCSs).  This work extends previous studies which correlate the properties of
the MMR with those of the host galaxy.
Our combined data sets show a mean trend for the metal-poor
(MP) subpopulation which corresponds to a scaling of heavy-element
abundance with cluster mass  \textit{Z $\sim$ M$^{0.30\pm0.05}$}.  No
trend is seen for the metal-rich (MR) subpopulation which has a scaling relation
that is consistent with zero.  We also find that the scaling exponent is independent
of the GCS specific frequency and host galaxy
luminosity, except perhaps for dwarf galaxies.

We present new photometry in \textit{(g',i')} obtained with Gemini/GMOS
for the globular cluster populations around the southern giant
ellipticals NGC 5193 and IC 4329.  Both galaxies have rich
cluster populations which show up as normal, bimodal sequences
in the colour-magnitude diagram.

We test the observed MMRs and argue that they are statistically real,
and not an artifact caused by the method we used.  We also argue
against asymmetric contamination causing the observed MMR as our mean
results are no different from other contamination-free studies.
Finally, we compare our method to the standard bimodal fitting
method (KMM or RMIX) and find our results are consistent.

Interpretation of these results is consistent with recent
models for globular cluster formation in which the MMR is determined by
GC self-enrichment during their brief formation period.
\end{abstract}


\section{Introduction and Background}
\label{introduction}

Colour bimodality in globular cluster systems (GCSs) has been observed
for nearly all types of galaxies, including small and large
galaxies, and elliptical and spiral galaxies (e.g., \citealt{2006ApJ...636...90H,
2006ApJ...653..193M, 2006ApJ...639..838P, 2006AJ....132.1593S, 2006AJ....132.2333S,
2007ApJ...668..209C, 2008ApJ...681.1233W}) and is now thought of as a
``universal'' characteristic that reflects their formation history.

This red/blue bimodality 
corresponds to a bimodality in metallicity.  Concern 
has been raised that the
split occurs because of a non-linear
colour-metallicity relation (e.g., \citealt{1994ApJS...95..107W,
  2006Sci...311.1129Y}); however, more direct metallicity
measurements, including IR colours \citep{2007ApJ...660L.109K}
and spectroscopy \citep{2006AJ....131..814B, 2007AJ....133.2015S, 2008MNRAS.tmp..502B},
argue against this.  In addition, the GCS for the Milky Way
is found to have a linear \textit{(B-I)}-metallicity relation (Figure
7 in \citealt{2006ApJ...636...90H}).  M31 also shows bimodality in
\textit{(U-V)$_0$}, \textit{(U-R)$_0$} and \textit{(V-K)$_0$} colours and
metallicity, through spectroscopic and photometric studies presented
in \cite{2000AJ....119..727B}.  Bimodality for M31 is not seen in other
colours, including \textit{(V-I)$_0$} - a colour often used previously
with the testing of bimodality.  \citeauthor{2000AJ....119..727B}
suggest, however, that photometric and reddening estimate errors blur away the
expected bimodality distribution in these colours (that are actually less
sensitive to metallicity).

Another concern is that age and metallicity are degenerate.  One
attempt to understand this issue was undertaken by
\cite{2005A&A...439..997P} who used Lick indices to look at the GCSs of seven early-type
galaxies.  They found that there are three age groups into which GCs
fall: two-thirds have an age $>$10 Gyr, one-third are between 5 and 10
Gyrs old, and a small fraction have an age $<$5 Gyrs.  The issue of age-metallicity degeneracy remains to be resolved, but the
spectroscopic data provide clear evidence that colour bimodality
transforms to metallicity bimodality.  If age alone could account for the GCS bimodality, a difference of
$\Delta (B-I)\approx$ 0.3 would correspond to age differences of 4.5 Gyr and
7.5 Gyr for constant metallicities of [Fe/H]=-2.25 and -1.5,
respectively, with the redder clusters being the older subpopulation \citep{2005MNRAS.362..799M}.

Remaining challenges to unravel are the presence of the two
distinct subpopulations and what these may mean for
galaxy formation theories.  The ideas can be grouped into three main categories:
\cite{1992ApJ...384...50A} suggest
gas-rich galaxy merging, where each progenitor galaxy
has its own blue, metal-poor (MP) GCS and then upon merging the two galaxies create a
red, metal-rich (MR) subpopulation.  In multi-phase 
dissipational collapse \citep{1997AJ....113.1652F} the GC
formation process is halted, for example, via cosmic reionization
after a galaxy forms its MP GCs \citep{2002MNRAS.333..383B, 2003egcs.conf..348S} and then resumes some time later to form
MR GCs.  Finally, \cite{1998ApJ...501..554C} suggest that an initially large galaxy has
its own MR GCs and then accretes MP GCs from satellite galaxies.
\cite{2003egcs.conf..317H} comments that these processes are not
mutually exclusive, and are likely to act in a number of combinations
for different circumstances. 

In addition to the bimodality, a correlation between luminosity and
mean colour -- that is, a mass-metallicity relation (MMR) -- is seen for
the blue subpopulation in at least some massive galaxies (e.g.,
\citealt{2006ApJ...636...90H, 2006ApJ...653..193M,
  2006AJ....132.1593S, 2006AJ....132.2333S}).  
This feature is also referred to as
the ``blue tilt''.  Drawing material from the previously
published literature, we list values in Table \ref{previousmmrs} of the
measured MMR slope along the blue GC sequence; this value, $p$, is the exponent in the scaling of heavy-element
abundance $Z$ with GC mass $M$, Z$\backsim$M$^p$.  
We will discuss the derivation of these slopes and their comparisons
in Sections 3 and 4.
Columns in Table 1 contain 1) the galaxy sample from which the blue-MMR value was
obtained; 2) the value of $p$ for the blue-MMR with errors 
where available; 3) the brightest
GC luminosity values that were used to make the fit, and 4) the
reference to the data source paper.

No obvious MMR along the 
\textit{red} sequence has been observed \citep{2006ApJ...636...90H,
  2006ApJ...653..193M, 2006AJ....132.1593S, 2008ApJ...681.1233W}.  However, \cite{2008ApJ...681.1233W} detected the red sequence
continuing to brighter magnitudes towards the ultra-compact dwarf (UCD)
galaxy range.  In addition, \cite{2006AJ....132.2333S} interestingly find no blue-MMR in M49 (NGC
4472).  If confirmed, this result will provide an additional
complexity to be explained by
galaxy and globular cluster
formation models.  
\cite{2006ApJ...653..193M} 
suggest that
environmental causes may produce this result because M49 is at the
centre of its own sub-cluster within
the Virgo Cluster.  This may be an example, they claim, of the
differences amongst galaxies with different accretion/merger histories
if MMRs are created through GC accretion from galaxies with different
masses.  Testing this hypothesis, they find that in general GC
accretion is unlikely to account for the entire slope of observed
MMRs, although it may contribute to or, in the case of M49, negate the trend.  

A few other galaxies with smaller GC populations have also been searched
for an MMR.  
Both red and blue slopes for the GC sequences in
NGC 1533 are consistent with zero
\citep{2007ApJ...671.1624D}.  This SB0 galaxy is neither in a
cluster nor is it a bright elliptical, which the authors imply would be
the necessary conditions to produce the self-enrichment that they
suggest causes the blue-MMR.  The Milky Way does not show any
evidence for a blue-MMR,
but this may be due to small-number statistics - as there are only
$\backsim$150 GCs \citep{1996AJ....112.1487H} - rather than the fact
that it is a spiral galaxy.  \cite{2006AJ....132.1593S} show that
blue-MMRs are not limited to ellipticals as they see a blue-MMR in the
Sombrero Galaxy (NGC 4594).  \citealt{2006AJ....132.1593S} classify
this galaxy as an Sa, although they note that \cite{2004AJ....127..302R} find it to be an
S0.  As with ellipticals, no MMR is seen for NGC 4594's red subpopulation.
A key point in assessing the presence or absence of a detectable slope
has proven to be the total cluster population in the galaxy (e.g., \citealt{2006ApJ...653..193M}):  if the MMR
slope is most noticeable for the high-luminosity end of the GC sequence
($M \gtrsim 10^6 L_{\odot}$), it will be unambiguously detectable only
in galaxies with thousands of clusters where the high$-L$ end is well
populated.  For most galaxies, and certainly for most spirals and dwarfs,
this upper end is thinly populated and the existence of the MMR is likely
to be not decidable.

\cite{2006AJ....132.1593S} claim that the blue-MMR is not due
to the addition of a group of another kind of objects at brighter
magnitudes (around \textit{M$_V$} $\backsim$ -11), but that the relation continues down to fainter
magnitudes - at least to the GC luminosity function turnover magnitude
of \textit{M$_V^{TOM}$} = -7.6 $\pm$ 0.06.  By contrast, \cite{2006ApJ...636...90H}
in their study of eight giant ellipticals 
find that the MMR slope is clearly nonzero only for the 
brightest blue GCs and that the blue sequence is essentially
vertical (i.e., no MMR) for $L \lesssim 6 \times 10^5 L_{\odot}$.
\cite{2006ApJ...653..193M} use a sample of several dozen Virgo galaxies,
binned into four groups by galaxy luminosity, to find that the
lowest-luminosity members (the dwarfs) show little or no MMR;
however, these are also the galaxies for which the high-luminosity
end of the GC sequence is least well-populated.

Though this is still a new feature of GCS systematics,
several model interpretations for the MMR have already been raised.
The most likely ones - both observationally and statistically - appear to have organized themselves around 
the early enrichment history of GCs during their formation, either
from their own internal self-enrichment or from their local
surrounding host gas clouds.  Working within
the hierarchical-merging picture of galaxy formation, 
\cite{2006ApJ...636...90H} suggest that the blue, MP GCs
form first while they are still within dwarf-sized and metal-poor host clouds.  
The more massive ones can enrich further during formation, thus giving
rise to an MMR.  By contrast,
the red-sequence GCs then form later and are slightly more
metal-enriched and red because they form in a deeper potential well when the galaxy
formation process has been completed.  \cite{2006AJ....132.2333S}
and \cite{2006ApJ...653..193M} 
favor self-enrichment as the cause of the blue MMR.
Quantitative models built on these ideas have been developed
by \citet{2008AJ....136.1828S} and \citet{2009arXiv0901.2302B}.

Although rare, Brightest Cluster Galaxies (BCGs) have uniquely large
samples of GCSs which
make them the best environments for studying MMRs.  In this paper we present newly
reduced GC photometric data
for two BCGs, NGC 5193 and IC 4329.  We follow with a new analysis
of the HST data for four more
BCGs and further combine them with the data from 
\cite{2006ApJ...636...90H} and \cite{2008ApJ...681.1233W}. 
We view our discussion in total as one more step towards understanding
this intriguing new correlation, which promises to uncover additional
clues towards GC formation and enrichment history.  Other papers by
the authors, both
on the observational and theoretical side, are in progress which will
help develop a more complete synthesis.

Section \ref{observations}
details the observations and data reduction of these two galaxies, and
is where we also obtain the
colour-magnitude diagrams (CMDs) and analyse the MMR.  We compare our work with other similar studies and discuss the results in Section
\ref{discussion}; we summarize our work in Section \ref{summary}.

\section{New Observations and Data Reduction}
\label{observations}

We first discuss our new observations for the giant galaxies
NGC 5193 and IC 4329.  
Some uncertainty exists about the status of NGC 5193 and the cluster
of galaxies of which it is a part.  At times it has been associated
with Abell 3560 (e.g. \citealt{1989ApJS...70....1A}).
However, its recession velocity and its projected position indicate that
it is \textit{neither} found at the centre of A3560 \textit{nor} is it
associated with that cluster.  The redshift of A3560 is $14470\pm123$
km s$^{-1}$
\citep{1999AJ....118.1131W}, whereas NGC 5193 and its companion (NGC
5193A) are at $3600$ km s$^{-1}$ \citep{1990AJ.....99.1709V}, and thus
NGC 5193 and NGC 5193A appear as
foreground objects.  NGC 5193 and NGC 5193A are located in 
Abell 3565  \citep{1999AJ....118.1131W, 2002A&A...396...65B}.  
Abell 3565, which has a velocity of $3586\pm45$ km s$^{-1}$ \citep{1999AJ....118.1131W} and a richness class of 
1 \citep{1989ApJS...70....1A}, is a cluster centred on 
IC 4296, the brightest galaxy in the cluster 
\citep{1990AJ.....99.1709V, 1999AJ....118.1131W}. 

IC 4329 is associated with Abell 3574 which is also known as 
Shapley 1346-30, K27, the IC 4329 group and Centaurus North 
(\citealt{1993Ap&SS.207...91P} and references therein).  IC 4329 also
has a richness class of 1 \citep{1989ApJS...70....1A} and a velocity
of 4808 $\pm$ 21 km s$^{-1}$ (the CMB reference frame; NASA Extragalacitic Database).  Abell 3574 is in turn a member of the Hydra-Centaurus Supercluster.

\begin{figure}[h]
  \begin{center}
  \includegraphics[width=105mm]{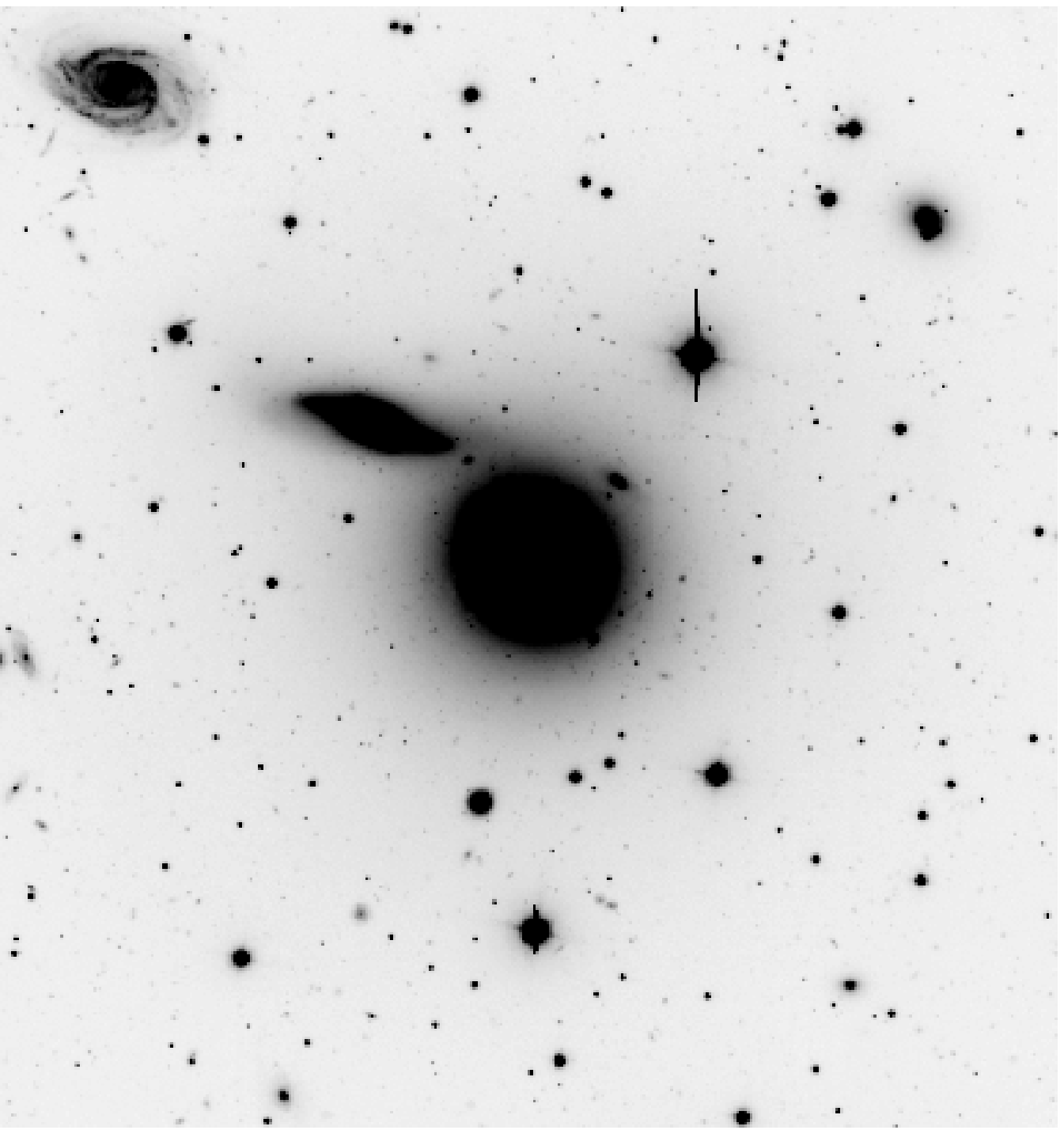}
  \includegraphics[width=105mm]{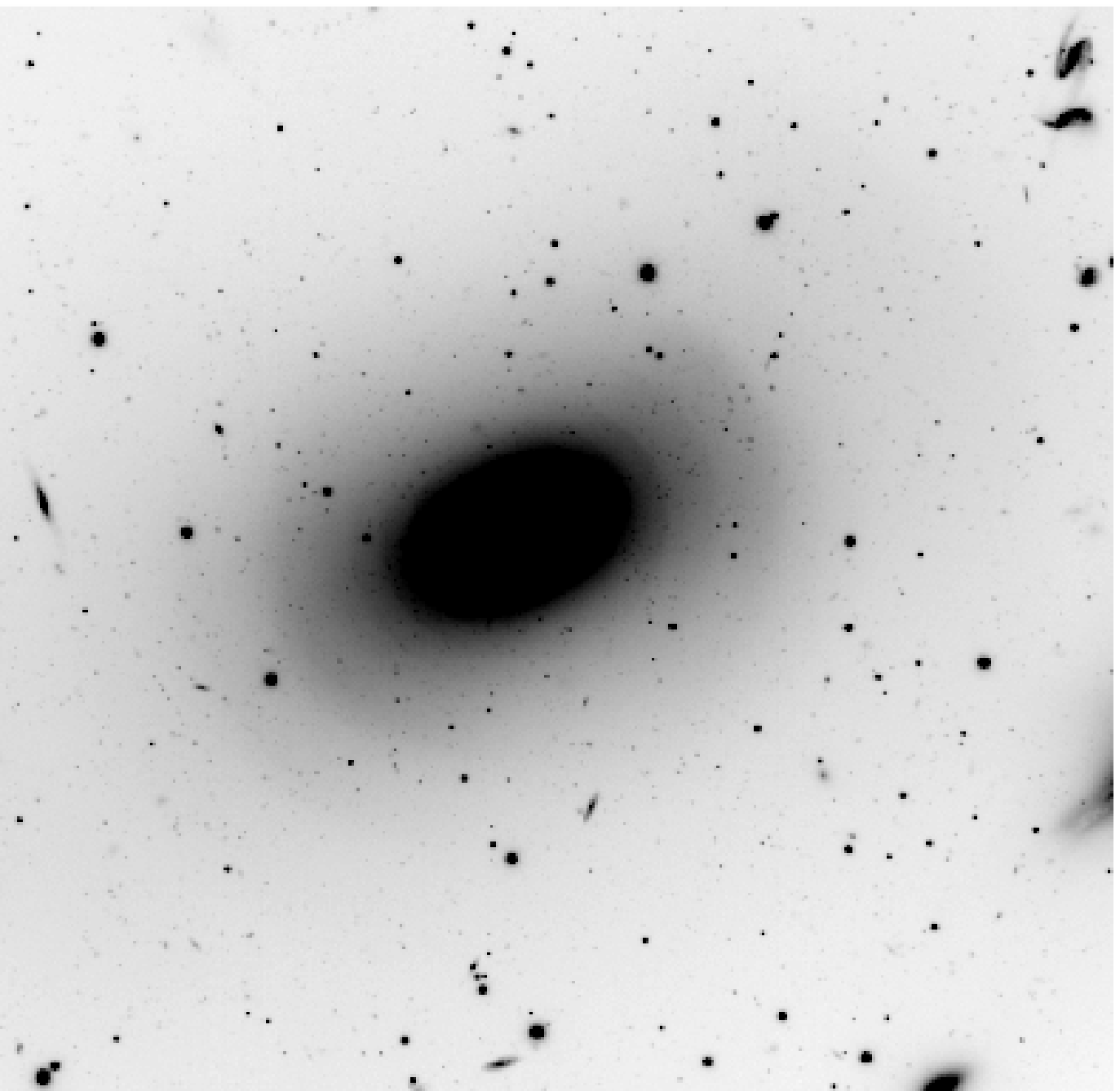}
  \end{center}
  \caption{NGC 5193 and IC 4329 Gemini GMOS $g'$-band images that have a
    5.5' x 5.5' field of view and an image scale of 0.146'' per
    pixel, and are from the data set GS-2006A-Q-24.}
  \label{51934329images}
\end{figure}

These two galaxies were observed with the Gemini-South
GMOS camera in $(g',i')$ on the nights of
2006 February 8 and March 23.  Images have a 5.5' x 5.5' field of
view, an image scale of 0.146'' per pixel, and come from the data set
GS-2006A-Q-24.  The two fields can be seen in Figure \ref{51934329images}.  
These images were taken during the same run as NGC 3311,
published in \cite{2008ApJ...681.1233W} whose data we re-analyse
here to compare with the data from NGC 5193 and IC 4329.
Raw exposure times were $13 \times 500$ sec in $g'$
and $12 \times 500$ sec in $i'$, similar to the data for NGC 3311.
We used the GEMINI/GMOS package within IRAF to
preprocess all the images for NGC 5193, IC 4329, and used standard-star
exposures taken during the run to calibrate the data.  Adopted values 
of redshift and luminosity are taken from NED.  The foreground extinction
values were evaluated by using the approximate 1/$\lambda$
relation and the extinctions in the UBVRI system as listed in NED.

NGC 5193 and IC 4329 are
at distances of 56.6 $\pm$ 3.8 Mpc and 72.4 
$\pm$ 4.6 Mpc (calculated from their values of \textit{cz}, in the CMB
reference frame, as in NED along with 
\textit{H$_0$} = 70 km s$^{-1}$).  For a typical GC
half-light radius of 3 pc, a GC at a distance of 60 Mpc would subtend
an angle of 0.021'' which is $\simeq$20 times smaller than the ground seeing
FWHM, and so GCs appear entirely star-like.  Thus, we can discount the
possibility suggested by \cite{2008AJ....136.1013K} that
the blue-MMR is only seen because of aperture-size biasses in high-resolution
data.  

\begin{figure}[h]
  \includegraphics[width=80mm]{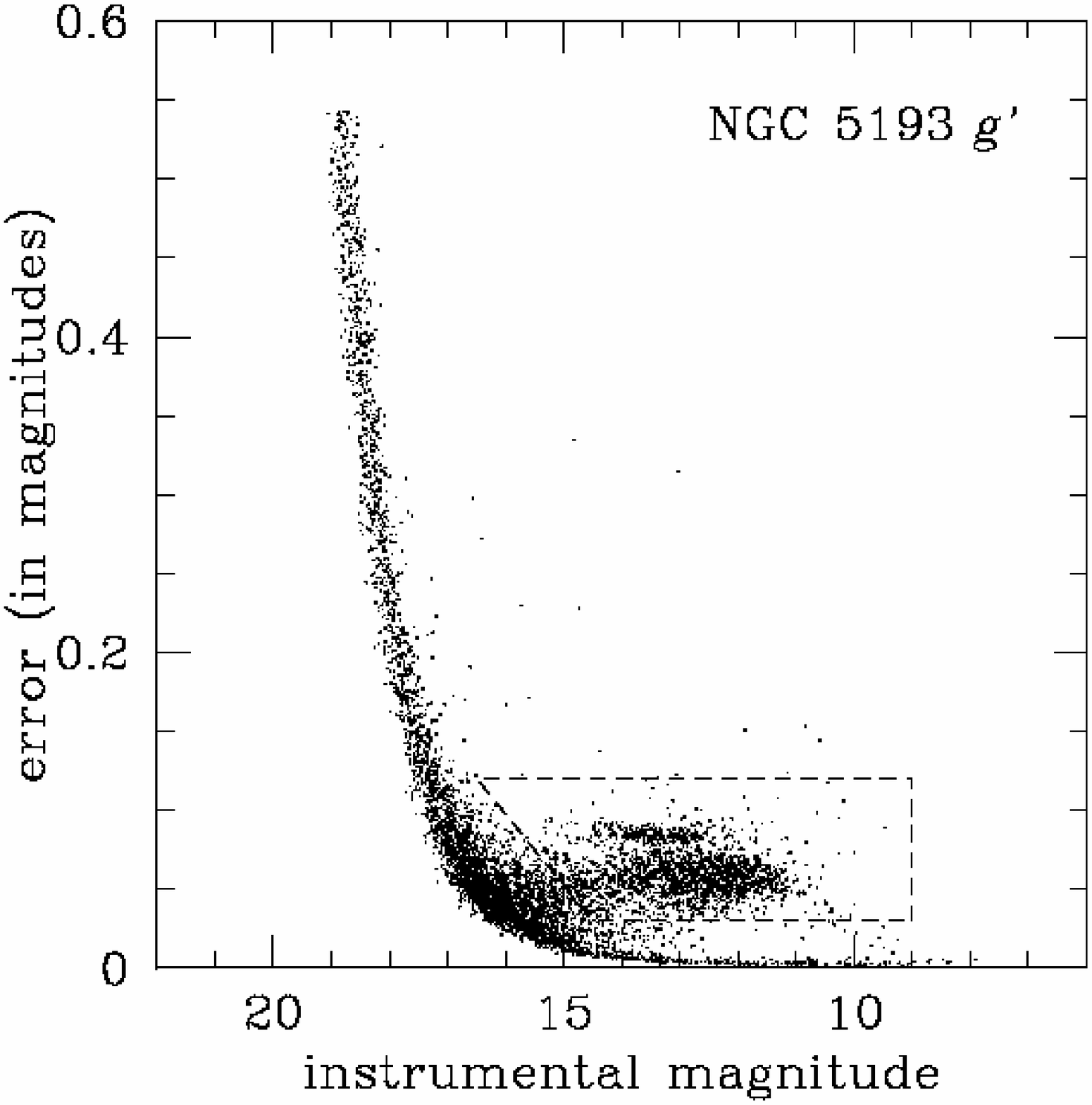}
  \includegraphics[width=80mm]{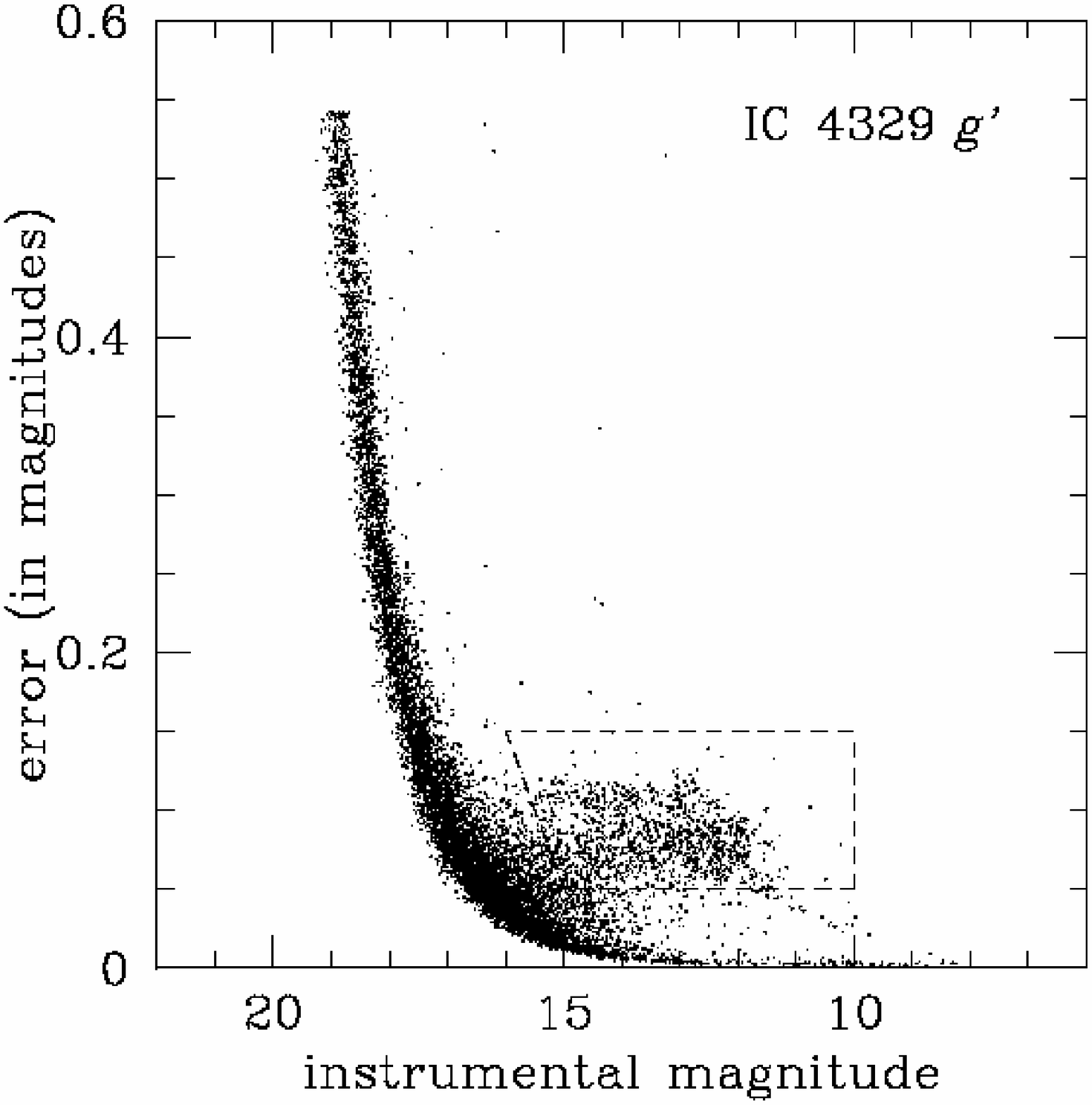}
  \caption{The NGC 5193 $g'$-band and IC 4329 $g'$-band data sets are
    shown with plots of instrumental magnitude versus error,
    and the false identifications within the
    dashed boxes.}
  \label{falseids}
\end{figure}

To prepare the images for photometry, we
median-subtracted the smoothed isophotal light,
allowing the faint star-like objects to be seen more easily
in to near the galaxy centres.
For the photometry, we used DAOPHOT (the standalone version 4) and
ALLSTAR in the same way as described in
\cite{2008ApJ...681.1233W}.
Two of the nine parameters that ALLSTAR returned - the magnitude of
the object and the associated error - proved useful to identify
false ``hits'' which were clearly associated with the higher background
noise near the brightest regions.  
These were removed from all later analysis.  
Examples of these false identifications are shown in the dashed boxes
in Figure \ref{falseids}.  

We used DAOMATCH to match the detected objects across the $g'$-
and $i'$-band images, with DAOMASTER refining the estimates.  Only x- and
y-shifts were allowed, leaving coordinate shifts accurate to within $\pm$ 0.1
pixels.  Finally, we also used IRAF's GEOMAP as a consistency check.
The final photometry files of matched objects measured in both $(g',i')$
comprise $1932$ objects for NGC 5193 and $3558$ for IC 4329.

\subsection{Calibration}
\label{calibration}
The photometric zeropoints were set from short exposures of
standard stars from \cite{1992AJ....104..340L}, with 
$(g',i')$ magnitudes from the catalog\footnote{The catalog can be
  found at
  http://www.physics.mcmaster.ca/$\backsim$harris/Databases.html} by
E. Wehner. The GMOS calibration images of these were obtained
on the nights of 2006 March 25, 26 and 27, the same nights as the
science images.  Because only a handful      
of standard-star exposures were made during the two nights, we view
these calibrations as strictly preliminary and note that we intend to
improve them in a future paper through a specially designed photometric
observing run.  Fortunately, as will be seen in the discussion below,
our analysis of the MMR and the bimodal sequences depend only on
the relative colours, not the absolute scale.  The peaks of blue and red subpopulations for many galaxies are virtually
constant at a value of $(g'-i')_0 \approx$ 0.8 and $(g'-i')_0 \approx$
1.05 (e.g., for M87 with extensive calibration to the SDSS system;
\citealt{Harris2009}).  Using this information, we estimate our colour
terms to be accurate for NGC 5193, and offset by $(g'-i')_0 \approx$
0.2 for IC 4329 and NGC 3311.

\subsection{Colour-Magnitude Diagrams}
\label{bimodality}

\begin{figure}[h]
  \begin{center}
  \includegraphics[width=75mm]{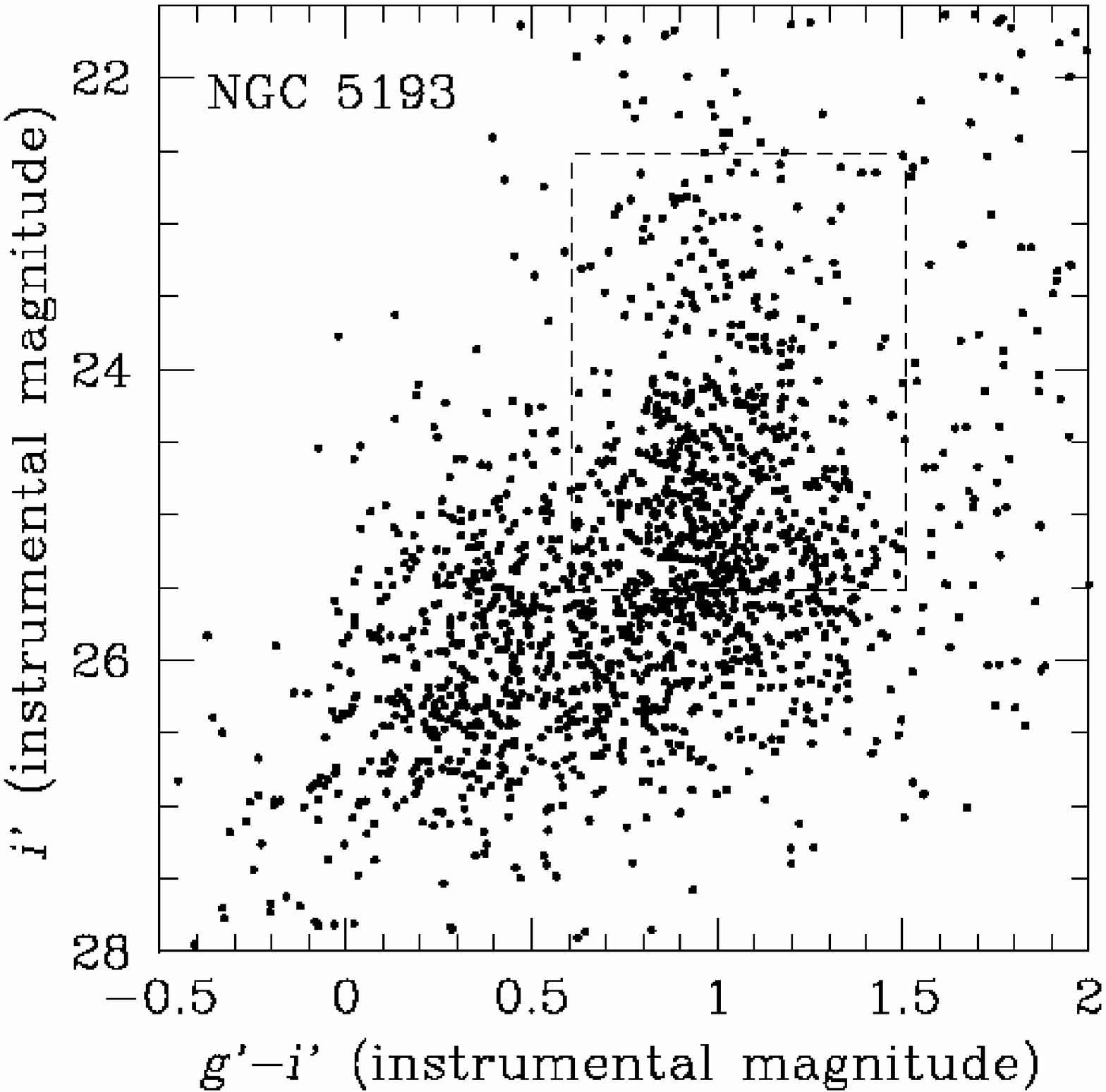}
  \includegraphics[width=75mm]{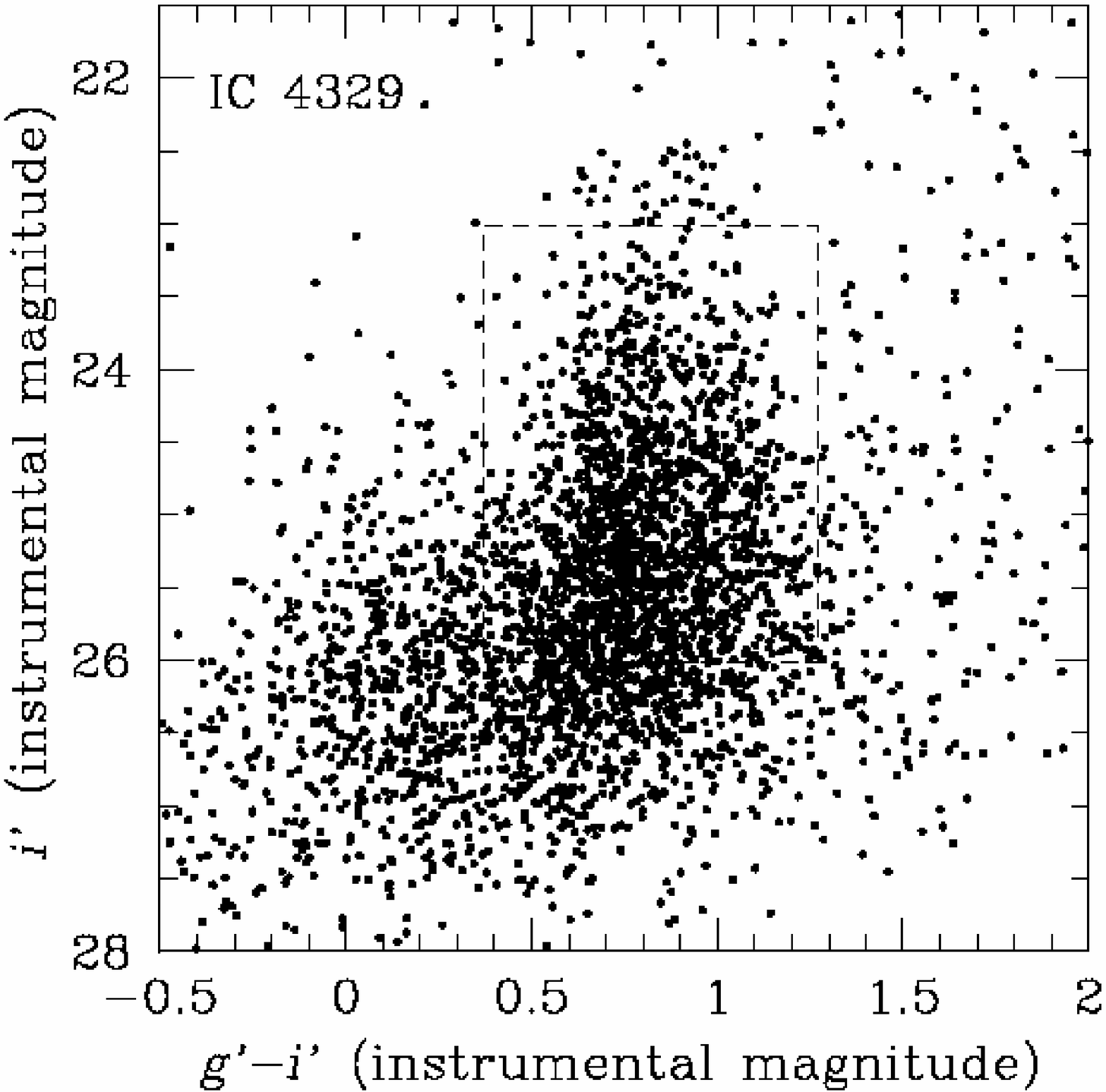}
  \includegraphics[width=75mm]{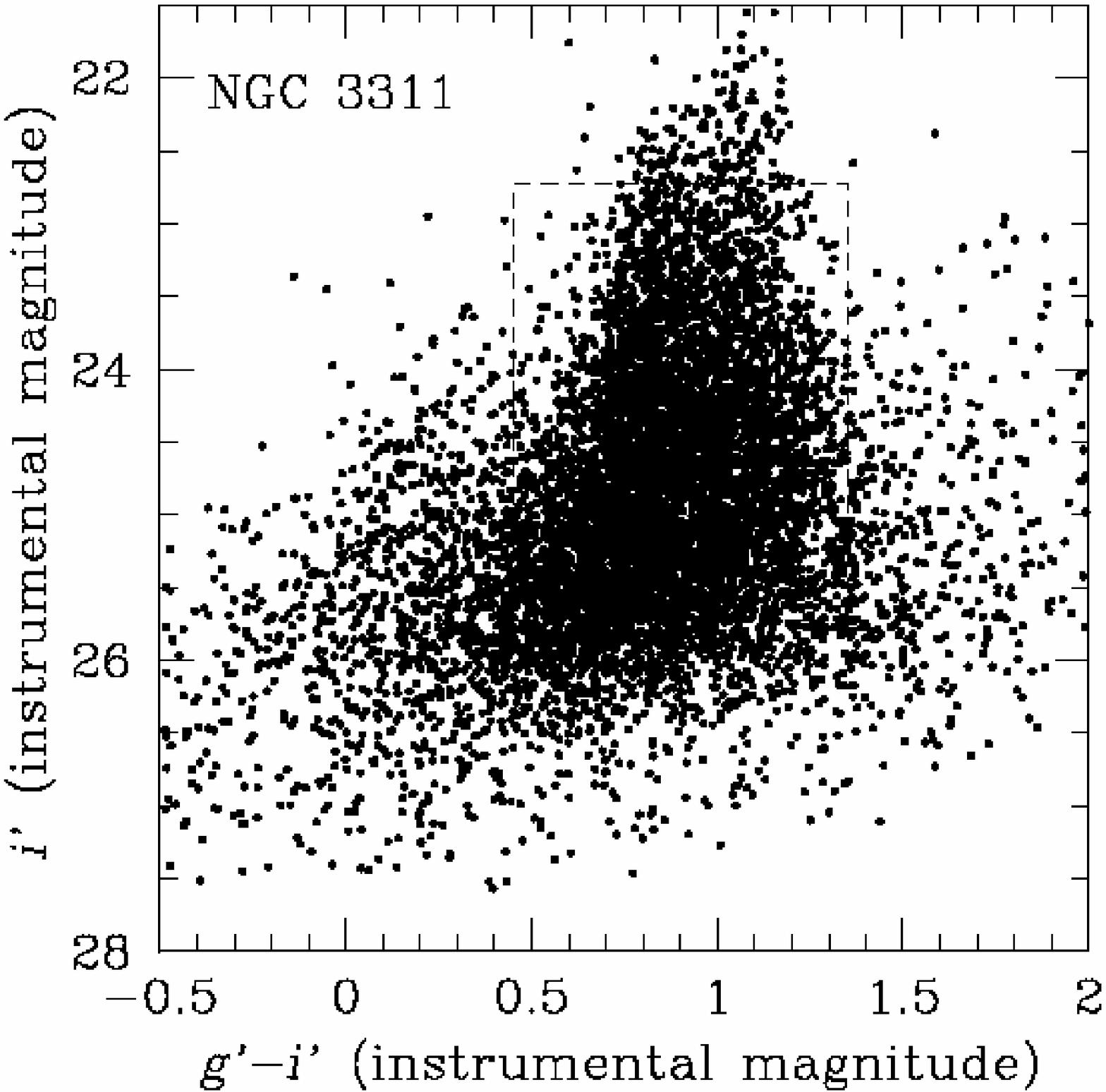}
  \end{center}
  \caption{Colour-magnitude diagrams in $(i', g'-i')$ for the
    two galaxies imaged in our Gemini GMOS program, NGC 5193
    and IC 4329, as well as that of NGC 3311 from \cite{2008ApJ...681.1233W} for comparison.  GC candidates are located within the dashed-line
    box.  As discussed in Sections \ref{bimodality} and \ref{mmr}, the
    brightest limit of the box was incrementally increased in 0.3
    magnitude steps.} 
  \label{2cmd}
\end{figure}

The colour-magnitude diagrams (CMDs) for the measured objects in
both fields are shown in Figure \ref{2cmd}, along with the CMD for NGC
3311 (shown for comparison).  In these,
the GCSs stand out as pairs of vertical sequences.  Areas enclosing
these vertical sequences were chosen as the areas of interest,
to eliminate obvious field contamination.  Bright limits were also
chosen so that a range of three magnitudes was initially
enclosed (while enclosing most of the GCSs, this also allowed the
authors to brighten the brightest limits in steps for the later
analysis discussed in Section \ref{mmr}).  Data points were selected from
an initial region enclosing the GCS vertical sequences within the
limits as follows:

\begin{itemize}
\item 0.5 $<$ $(g'-i')_0$ $<$ 1.4 and -11.35 $<$ M$_{i'}$ $<$ -8.35 for NGC 5193, 
\item 0.25 $<$ $(g'-i')_0$ $<$ 1.25 and -11.4 $<$ M$_{i'}$ $<$ -8.4
  for IC 4329 and
\item 0.3 $<$ $(g'-i')_0$ $<$ 1.2 and -11.3 $<$ M$_{i'}$ $<$ -8.3 for
  NGC 3311\footnote{The colour range for NGC 3311 is the same as
    \cite{2008ApJ...681.1233W}, the fainter limit is also similar, and
  the brighter limit was fixed to form our standard initial enclosure
  of three magnitudes above the fainter limit.}. 
\end{itemize}
\noindent We note that these selection regions are not identical in
colour range, but note again that they were chosen with the aim of
enclosing the GCS while minimizing field contamination.  As also noted above, this is likely to be
\textit{not} because of
intrinsic differences between the galaxies, but rather
because of remaining errors in the photometric zeropoints as the
different galaxies were observed on different nights; but again,
the bulk of our later analysis depends only on the
relative colours and luminosities.  The number of data points within this initial
selection region is shown in column (7) of Table \ref{galaxydetails}.
We choose to use \textit{M$_{i'}$} as the GC luminosity parameter
(as opposed to \textit{M$_{g'}$}) because the values of \textit{M$_{i'}$} most
closely resemble the bolometric luminosity,
and because plotting against the redder colour more accurately
portrays the MMR (\citealt{2006ApJ...653..193M}; 
see also our Section \ref{comparisonstootherdata}).  The faint-end limits of \textit{M$_{i'}$}$>$-8.35 and -8.4
were estimated visually and adopted to minimize field contamination.  The bright-end limits of
\textit{M$_{i'}$}$<$-11.4 and -11.35 were only initial and were later
increased in steps (see below) to see what effect, if
any, this would have on the deduced MMRs. 

In Figure \ref{radialdist}, we show the radial distribution of the GC candidates
around each galaxy (see Appendix for a more detailed description
of these diagrams, including calculation of the GCS specific
frequencies).  As the CMDs also indicate,
the GCS around NGC 5193 is the less populous of the two, and
is also the more centrally concentrated with a steeper power-law
fit.

The first step in the analysis is to test the colour distribution
of the GCs for clear presence (or absence) of bimodality.  To do this,
we used the multimodal statistical fitting package RMIX \footnote{RMIX
is publicly available at
http://www.math.mcmaster.ca/peter/mix/mix.html} in the same way as
described more fully in \cite{2008ApJ...681.1233W}.  
We carried out one-, two- and three-Gaussian fits with 
the sample divided into colour bins of $(g'-i')$ = 0.045,
and for comparison we included the previously published NGC 3311 data
as well.
A normal chi-squared test shows that two modes are strongly preferred
over only one, but three modes produced no further improvement with
either the hetero- or homoscedastic fits.  A homoscedastic fit forces the dispersions of both subpopulations to
be the same, whereas a heteroscedastic fit allows them to be
independent.  We therefore argue that 
the two-Gaussian fit is the optimal solution, indicating that
these galaxies are basically similar to other giant ellipticals.  The
RMIX fits are shown in Figure \ref{rmix} and all parameters for these fits
are shown in Table \ref{rmixparams}.  Columns contain (1) the
target's NGC or IC number, (2) the
proportion of the subpopulation, (3) the error on the proportion, (4)
the mean value of the Gaussian fitted to the subpopulation (i.e., the
peak colour value),
(5) the error on the mean, (6) the sigma, or distribution value, (7)
the error on sigma (not applicable for the second subpopulation
because the values of sigma for both subpopulations was forced to be
equal), (8) the degrees of freedom in the fit, (9) the $\chi^2$ value, and (10)
the reduced $\chi^2$ value, found by dividing $\chi^2$ by the degrees
of freedom.  We argue the use of homo- versus heteroscedastic fits for several
reasons.  Significant overlap between the two subpopulations for all
three galaxies - NGC 5193, IC 4329 and NGC 3311 - causes the peaks in
the distribution to become obscured.  Allowing a
larger number of free parameters in such a scenario makes an RMIX fit harder to
converge.  When we allow a heteroscedastic fit to the subpopulations
of IC 4329, we find that the sigma values are equal within errors.
\cite{2008ApJ...681.1233W} obtain a similar result for NGC 3311.  However, for NGC 5193
we find that a heteroscedastic fit has discrepancies of $\sigma
(g'-i')$ = 0.16$\pm$0.01
 and 0.10$\pm$0.2 for the blue and red
 subpopulations, respectively.  With heteroscedastic fits, we also
 find that the proportion of GCs within each subpopulation becomes
 more unrealistic, especially for NGC 5193: 0.86$\pm$0.06 and 0.14$\pm$0.06 for the blue and red
 subpopulations in NGC 5193, respectively, 0.61$\pm$0.14 and 0.39$\pm$0.14 in IC
 4329, and 0.78$\pm$0.04 and 0.22$\pm$0.04 in NGC 3311.  As we were
 concerned that these above effects were amplified by NGC 5193 GCS's
 low numbers, we forced all the fits to be homoscedastic.  Even though the reduced chi-squared value for the two-Gaussian fit on
the NGC 3311 GCS is the lowest of the different fits tried, we note
that it is still unusually high.  Various smaller bin sizes were
tried in the two-Gaussian fit, but the proportions and peak values of
the subpopulations remained the same.  

The peak values of the red and blue subpopulations for the three Gemini
galaxies (including comparison with NGC 3311) are not the same,
again because of residual zeropoint errors as noted above.
However, it is important to note that the \textit{differences} 
between the colours of the blue and red peaks
$\Delta(g'-i')_{(red-blue)}$, are the same among all three
galaxies, indicating that their metallicity differences are similar.

\subsection{Analysis of the Mass-Metallicity Relation}
\label{mmr}

Quite simply, a mass-metallicity relation for either the blue or red GC sequences
will exist if the mean colour changes \textit{systematically} with luminosity.
The two techniques that have previously been used 
to find it are either
to assume a linear relation between mean colour and absolute magnitude
and solve for the slope \citep{2006AJ....132.2333S, 
2006AJ....132.1593S, 2007ApJ...671.1624D, 2008ApJ...681.1233W}; or,
to bin the data in magnitude steps and find the mean colour of
each sequence in each bin \citep{2006ApJ...636...90H, 2006ApJ...653..193M,
2008ApJ...681.1233W, Harris2008}.  In this paper, we use the former as
the main method and compare to the latter.

The data set for each GCS was sorted into 0.3 magnitude bins in
$M_{i'}$ so that enough bins along the sequence were generated while
also maintaining enough GCs per bin to be statistically significant.
Then to reduce field-star contamination further, we applied an
approximate representation of the Chauvenet
criterion \citep{1961Parratt} to each bin.  The criterion's purpose is to reduce
the biasing effect of outlying points on values for the mean and
standard deviation.  We found the mean colour and its uncertainty 
$\sigma_{mean}$, for each bin and discarded outlying points if their
colours fell farther from the bin mean than z$\sigma_{mean}$ , where

\begin{equation}
\label{chauvenet}
z\simeq0.85(log_{10} n)+1.12 
\end{equation}

\noindent and n is the number of points in the bin.

To get colour-magnitude slopes (i.e., $\Delta (g'-i')/\Delta M_{i'}$), we
first split the data points into red and blue sides with a line at one
colour value, although the value of this
initial dividing line varies from GCS to GCS as discussed below.  Figure \ref{initialsplit} shows the published data in NGC 4696, which
we use as an example for illustrative purposes.  The middle dashed
line in this figure represents the initial dividing line.  The value of the initial dividing line for each GCS is chosen to be at the ``dip''
colour value (\citealt{2006ApJ...639...95P, 2008ApJ...681..197P};
i.e., the colour where a GC has equal probability of being red or
blue).  \cite{2008ApJ...681..197P} note that although the mean colours
of subpopulations may vary with galaxy luminosity, the
dip colour is nearly the same.  As our data sets have yet to be
well-calibrated we can see that our dip colours are not all the same:
IC 4329 and NGC 3311 have very similar dip colours of
\textit{(g-i)$_0$}=0.78 and 0.80, respectively, whereas NGC 5193's dip
colour is \textit{(g-i)$_0$}=1.02.  However, knowing that the dip
colour values \textit{should} be constant, we choose the dip
colour of the GCS as the initial dividing line.  We then binned the 
two data subsets in 0.3 magnitude bins, finding a mean for each bin.  
Finally, we calculated a weighted least-squares (WLSQ) linear fit of
bin magnitude versus bin colour for each subset (the upper and
lower dashed lines in Figure \ref{initialsplit}).

A line (the middle solid line in Figure \ref{initialsplit}) exactly
half-way between these two initial fits was used to split the
Chauvenet-excluded data set a second time, and the process was repeated.  This
second splitting was to compensate for any inequality in the
proportions of the two subpopulations (see Section \ref{comparingdifferentmethods}).  The upper and lower solid
lines in Figure \ref{initialsplit} show the final WLSQ fits.  
Figure \ref{51934329mmrs} shows the CMDs and MMRs for NGC 5193 and IC
4329 (with only 11 and 12 bins, respectively), where we also added the one-sigma error bars of the WLSQ fit to the
adopted linear solutions.  If we were to eliminate the brightest bin
in the red slope of IC 4329, we can see that within errors the fitted slope would
immediately drop to zero.  This outlying bin is excluded on similar
grounds to the Chauvenet criteria, as was a similarly outlying point
excluded in NGC 5322.

As noted above, the upper, brighter limit of the
selection region was incrementally increased 
in 0.3 magnitude steps to explore what effect -
if any - it had on the colour-magnitude slope. We used three methods
to do this.  For each subpopulation in each GCS in our first method, we started with a minimum of ten
bins for which we calculated a slope.  Bins at brighter magnitudes
were added until there were
no more GCs to bin.  With each new bin added a new slope was
calculated.  This gave us up to five slopes per subpopulation per
GCS.  NGC 5193 had so few points that
only the slopes for 10 bins and 11 bins could be measured (i.e., only
one bin was added to the bright end).  The two
blue $\Delta(B-I)_0/\Delta M_I$ \footnote{$\Delta (g'-i')/\Delta
  M_{i'}$ slopes were converted to $\Delta(B-I)_0/\Delta M_I$ slopes
  via the relations \citep{2005MNRAS.362..799M} in Section \ref{comparisonstootherdata} to allow
  comparison to other data sets.  Errors are propagated through
  the equations.} slopes were the same within errors
(-0.05$\pm$0.03 and -0.05$\pm$0.02),
as were the red (-0.01$\pm$0.05 and -0.03$\pm$0.04).  For IC
4329, slopes for 10-13 bins were calculated (i.e., three bins were
successively added to the bright end).  The IC 4329 blue slopes varied
between $\Delta(B-I)_0/\Delta M_I$=-0.11$\pm$0.02 and
-0.05$\pm$0.01, and we found that the red varied between
$\Delta(B-I)_0/\Delta M_I$=-0.07$\pm$0.03 and 0.05$\pm$0.02.  The values of all blue and red
slopes are shown in Table \ref{redblueslopestable}.  The IC 4329 slopes for 12
and 13 bins are shown with and without the outlying point.

The second method used a constant number of ten bins.  As a brighter
bin was added, the faintest bin was dropped and a new slope was
calculated.  Again, this gave up to five slopes per subpopulation per
GCS.  To look at any possible curvature of the
MMR we used the IRAF/POLYFIT routine for our third and final
method with a polynomial of order two.  We input both the minimum number of ten bins, and also the maximum
number of up to fourteen bins in the routine and compared the
results.  

With the first two methods, we
saw no consistent change on the MMR slopes as brighter bins were
included.  With the final method, most slopes had an insignificant
second-order component.  The remaining few only had a marginal
second-order component.  We conclude that a linear slope for the MMR
is a reasonable assumption for our sample of galaxies, and that the
particular choice of bins (luminosity range) has no major effects on
the deduced slope.

To compare all slopes, we used the slope with the maximum number of
bins, but excluding the faintest and brightest bins.  This was done to reduce the effects of field
contamination (at the faint end) and small numbers (at the bright end).

\section{Comparison and Discussion}
\label{discussion}

Part of the purpose of our study is to search further
for any trends in the slope of the MMR with such factors
as host galaxy luminosity or environment.
To our MMR solutions for NGC 5193 and IC 4329,
we now add similar results from other BCGs or giant ellipticals.
Previously published ones include the eight in
\cite{2006ApJ...636...90H} and NGC 3311 in \cite{2008ApJ...681.1233W}.
To these, we add a 
further four BCGs - NGCs 7626, 708 and 7014, and IC 4296
\citep{Harris2008} - observed with HST in $(B,I)$, similar to
those in \cite{2006ApJ...636...90H}.  For the total of 15 BCGs listed above, 
Table \ref{galaxydetails} lists in successive 
columns (1) the target's NGC or IC number,
(2) the galaxy group or cluster, (3) the redshift \textit{cz} of the galaxy,
(4) the galaxy luminosity, \textit{M{$_V^T$}}, (5) the foreground
reddening, (6) the apparent distance modulus (either SBF-based
distances from \cite{2001ApJ...546..681T} were used, as is the case
for the galaxies marked with (i), or these values were calculated from
radial velocities using an
adopted value of \textit{H}$_0$ = km s$^{-1}$ Mpc$^{-1}$), (7) the number of
data points within the initial selection region (see Section \ref{bimodality} for details of the selection
regions), (8) the total number of GCs estimated to be in the system,
and (9) the specific frequency, S$_N$. 

\cite{2006ApJ...653..193M} conducted an MMR study for 79 early-type 
galaxies with HST/ACS photometry in $(g_{475}-z_{850})$ as part of the 
Virgo Cluster Survey.  Though most of these galaxies are dwarfs
or ones with small globular cluster populations in which no individual
MMR solutions can be determined, they combined their sample 
into four bins grouped by galaxy luminosity.  In each bin the total
GC sample is then large enough for any cumulative MMR slope to be found
with some confidence.  These four bins are compared with the BCG data
sets.  

\subsection{Comparisons to Other Data}
\label{comparisonstootherdata}

To put all these studies onto a common system for comparison
requires converting three different colour indices:
$(g_{475}-z_{850})$, $(B-I)$, and
$(g'-i')$.  In each case we convert $\Delta(colour)/\Delta(mag)$
into a scaling exponent $p = \Delta \rm{log} Z/\Delta \rm{log} M$.
There is currently no established
relation between $(g'-i')$ and metallicity for GCs, so we first converted
$(g'-i')$ to $(B-I)$, although we note that this double
conversion introduces larger external uncertainties.  With single stellar population models from \cite{2005MNRAS.362..799M} we calculated the linear
colour conversion: 

\begin{equation}
\label{fromgitoBI}
\Delta (B-I) = \frac{\Delta(g'-i')}{(0.53\pm0.12)}.
\end{equation}

\noindent To convert to metallicity from \textit{(B-I)} colours, we used the
linear relation as in \cite{2006ApJ...636...90H}, which is based on
colours and metallicities for Galactic GCs:

\begin{equation}
\label{Harrisconversion}
\frac{\Delta(B-I)_0}{\Delta[\textrm{Fe/H}]} = 0.375 \pm 0.049
\end{equation}

\noindent \cite{2006ApJ...653..193M} obtain their $(g_{475}-z_{850})$ colour-metallicity conversion
from \cite{2006ApJ...639...95P}, who have a piecewise relation that is
based not only on Galactic GCs but also GCs in M87 and NGC 4472:

\begin{equation}
\label{Pengconversionblue}
\frac{\Delta(g_{475}-z_{850})}{\Delta[\textrm{Fe/H}]} = 0.195 \pm
0.025 \hspace{1cm}[0.70<(g_{475}-z_{850})\leqslant1.05]
\end{equation}

\begin{equation}
\label{Pengconversionred}
\frac{\Delta(g_{475}-z_{850})}{\Delta[\textrm{Fe/H}]} = 0.546 \pm
0.069 \hspace{1cm}[1.05<(g_{475}-z_{850})<1.45]
\end{equation}

\noindent We note that \cite{2006ApJ...639...95P} actually mean log Z rather than
[Fe/H].  These two quantities are different by a factor of
[$\alpha$/Fe], which we assume to be constant so that the $slope$ remains the same when
converting between colour and metallicity.

The compilation of results for this entire extended sample of galaxies
is shown in 
Figure \ref{comparisonzoomedout}, as a graph of scaling slope $p$
against host galaxy luminosity.  Here, we show the 
MMR for each galaxy for both the blue and the red subpopulations.  
For the \cite{2006ApJ...653..193M} data, each galaxy has two data
points plotted.  The data set represented by open triangles is
$(g_{475}-z_{850})$ versus $g_{475}$-magnitude, while the data set
represented by solid triangles is
$(g_{475}-z_{850})$ versus $z_{850}$-magnitude.  \citeauthor{2006ApJ...653..193M} argue
that the $(M_{z_{850}},(g_{475}-z_{850}))$ relation is to be preferred, because
cross-contamination biasses between the two modes affect the
$(M_{g_{475}},(g_{475}-z_{850}))$ relation more strongly.  
We consider only the $(M_{z_{850}},(g_{475}-z_{850}))$ data points (solid triangles) in our discussion. 

For galaxies brighter than $M_V^T \simeq -19$, we find that the blue
slopes have a mean $p-$value
of $p=0.30\pm0.05$.  For the red slopes, we find
$p=-0.1\pm0.1$, which is consistent with zero.  $P$-values of blue and
red slopes are shown in Figure \ref{comparisonzoomedout}.  The dwarf
galaxies in the faintest bin of \cite{2006ApJ...653..193M} have
$p=0.11\pm0.18$ for the blue-MMR.  However, we note that this lower
value may be because there are fewer high-luminosity GCs in that bin - even after adding all the dwarfs together.  Therefore, if the MMR is
more prominent at higher parent galaxy luminosities, it would affect the dwarf/giant
comparison.  We find a consistent $p-$value roughly independent of the
individual GC
luminosity range, although at increasingly fainter levels down along
the GC sequences the
determination becomes less certain because of increasing photometric
scatter and field contamination.

Note that our error of $\Delta p$=0.05 is an
$internal$ error on the slope and does not include the transformation errors from $(B-I)_0$ to metallicity or from $(g'-i')$ to metallicity,
which are approximately $\Delta p \approx$0.02 and 0.2, respectively.
These additional $external$ uncertainties are the inevitable result of
combining different photometric systems, but as far as we can
determine, they do not affect our overall conclusion that $p \approx$
constant for the more luminous galaxies.  Further, more strictly
homogeneous surveys would remove this lingering uncertainty.

\subsection{Comparing Different Methods}
\label{comparingdifferentmethods}
We now test whether or not the line splitting method
we used above
may have actually introduced the observed MMR.  Suppose we have
a situation where the mean colours of both red and blue modes
stay constant with luminosity, but where the numbers of GCs in
the two modes might be different.  Suppose also (realistically)
that the random measurement uncertainties increase with
magnitude, so that the two modes would overlap more and more
as we go to fainter magnitudes.  If the two
subpopulations contained \textit{equal} numbers of GCs, MMRs would not be seen
(i.e., they would be consistent with zero) because similar numbers
of clusters in each mode would contaminate the other mode.  However, if the blue
subpopulation dominated, and the same constant dividing line was used,
a positive offset for both subpopulations would be seen:
more blue clusters would spill into the red side than vice versa,
so the measured mean colour of both modes would be too blue at
the faint end of the sample.  In this case, we would conclude
that the mean colours of both modes became redder at brighter magnitudes.
Conversely, if
the red subpopulation dominated, a negative offset for both
subpopulations would be seen.  For this reason, we used the second iteration
of the line splitting to counter this effect.

We do not see this contamination-based effect happening in any
of the solutions for our sample galaxies.  For the blue mode,
the $p-$values are consistently positive, whereas for the red mode
the slopes are slightly negative and consistent with zero.
In particular, for the red-dominated GCSs (NGCs 1407, 5322,
7049, 3348, 4696ap, 5557 and 7626, and IC 4296) the red-mode slopes are all
slightly negative or close to zero, which is consistent
with the above argument.  However, in these same GCSs the blue-mode MMRs
have an average value around p $\backsim$ 0.30, which is \textit{not}
consistent with the argument and therefore provides some support 
for the analysis method.  

Another and different factor, which we cannot assess as
completely here, is the amount of field contamination.
Although we selected restricted intervals in colour to minimize 
the contamination, it is notably worse at 
fainter magnitudes and if  the contamination is asymmetric 
in colour, it could bias the RMIX fits. An argument against
such an effect, however, is that the \cite{2006ApJ...653..193M} Virgo sample
is contamination-free (the GCs were identified individually)
and their mean results are no different from ours at the same
galaxy luminosities.  In addition, the different galaxies in our
sample have different degrees of field contamination (cf. the individual
source papers) but yield very similar MMR slopes.

We have also compared our results with methods using the KMM bimodal
fitting code, including six galaxies from
\cite{2006ApJ...636...90H}, and for the three Gemini/GMOS galaxies by conducting
RMIX fits.  (Only six of the eight \citeauthor{2006ApJ...636...90H}
galaxies are used because two galaxies only have one bin per
subpopulation; each bin contains 200 points.)  The Gemini data sets were divided into 0.5
magnitude bins, with the brightest two bins in NGC 5193 and IC 4329
not being included in the fit because of the very low numbers of GCs
there.  Comparisons were made between RMIX and KMM, as shown in Table \ref{slopevskmmtable} and Figure
\ref{comparingbrslopes}.  Six of the nine blue-MMRs and a different six of the nine red-MMRs are
consistent within errors for both methods, further supporting the line
splitting method and showing that the codes are consistent.

We find a blue-MMR index at a constant value of 0.30$\pm$0.05.  However, two
galaxies are clearly outside that range: NGC 3311, which has a much
higher index, and IC 4296 which has an index that within the errors is
consistent with zero.  We also recall that M49 (NGC 4472), included in
the brightest bin of the \cite{2006ApJ...653..193M} data set, also does
not show a blue-MMR.

\subsection{No Correlation with Specific Frequency}
\label{pointstonote}
The fifteen GCSs in Table \ref{galaxydetails} can be split into 
two groups: sparse GCSs, which
we define in a similar manner to \cite{2006ApJ...636...90H} 
to have specific frequency, \textit{S$_N$}$\leq$5; these include NGCs
1407, 5322, 3348, 3268, 5557, 7626 and 5193, and ICs 4296 and 4329; and rich GCSs, 
defined to have \textit{S$_N$}$>$5; these include NGCs 7049, 3258, 4696, 708, 7014 and 3311 (see the Appendix in Section
\ref{appendix} for $S_N$ calculations).  We find no significant difference
in the average MMR slope ($p-$value) between these two groups (see
Figure \ref{slopessn}), although we note that NGC 3311 is a
significant outlier.  The lack of correlation of $p$ with $S_N$ is consistent with the
interpretation that the source of the MMR is one more local to the GCs
themselves (such as self-enrichment) rather than one driven by the
large-scale GC formation efficiency.

\section{Summary}
\label{summary}

We discuss the photometry for globular cluster systems in
fifteen BCGs and use these to derive the MMRs for both the red
and blue globular cluster subpopulations.  Twelve data sets were from
the HST, and a
further three data sets were obtained from Gemini/GMOS with
$(g',i')$ photometry.  The eight ACS/WFC data sets and one 
Gemini data set were previously analysed in \cite{2006ApJ...636...90H}
and \cite{2008ApJ...681.1233W}, respectively.  For the other
two GMOS datasets (NGC 5193 and IC 4329) we present the photometry
and its results for the first time.

Colour-magnitude slopes were found via a ``line splitting'' method,
and converted to an MMR in heavy-element abundance versus mass
($Z \sim M^p$) so that comparisons between the HST and Gemini
data, and with the Virgo Cluster Survey galaxy sample
\citep{2006ApJ...653..193M} data, could be made.  We use the
composite results to construct a plot of the scaling exponents -
$p(blue)$ and $p(red)$ - versus host galaxy luminosity and also versus
specific frequency.

Our work confirms the existence of a blue-MMR with \textit{Z$\backsim$M$^{0.30\pm0.05}$},
independent of host galaxy luminosity for any systems brighter than
$M_V^T \simeq -19$.  Individual galaxy-to-galaxy differences may still exist (with occasional anomalies
such as NGC 4472, whose lack of an MMR still remains to be explained) but are
still hard to determine unambiguously.
For the dwarf galaxies fainter than $M_V^T \sim -19$, the blue-MMR slope is consistent with
zero, though we raise the question that this might be due to the smaller number
of GCs at high luminosities, making the MMR less noticeable.  Whether or not
the MMR continues to lower GC luminosities with the same slope remains an open
question, because of the increasing effects of field contamination and photometric
measurement scatter.

Within the measurement uncertainties, the red-sequence
MMR is consistent with a line that includes zero 
(\textit{Z} $\backsim$ \textit{M}$^{-0.1\pm0.1}$).  We also find no
correlation between the $p$-values of slopes and specific frequencies.

We have tested our line splitting method numerically to confirm
that that nature of the method does not itself artificially
introduce an MMR.  In general we find clear agreement between
the line splitting method and multimodal histogram fitting methods
employing RMIX and KMM. 

The results of this paper agree partially but not completely with
\cite{2006ApJ...653..193M}, who show that for lower
host-galaxy luminosities, i.e., dwarf galaxies, the blue-MMR
is not the same as those for mid-range host-galaxy luminosities.  Here, we
add many more galaxies to the brightest range and find similar
relations to those of the mid-range.  We note that NGCs 3311 and 4296 lie
offset from the rest of the galaxies for the blue-MMR in Figure
\ref{comparisonzoomedout}.  Only the dwarf ellipticals stand
out as having no clear MMR amongst their globular clusters.  This suggests that similar
GC formation processes occurred in BCGs as occurred in typical early-type galaxies.

The clear existence of a blue-sequence MMR is, in general, consistent with
recent models in which the higher mean metallicities within more massive GCs
are due to self-enrichment within the proto-GC: they form in more
massive potential wells which can hold in a higher fraction of 
supernova-driven enrichment (e.g., \citealt{2009arXiv0901.2302B}). Within the same interpretive framework, the lack of a similar trend along the red GC
sequence is expected if their mean metallicity is due to a higher level
of pre-enrichment, above which any later self-enrichment would be less
important.  The independence of $p$-values versus both host galaxy luminosity
and specific frequency may favour interpretations where the MMR is due
to local events in and near where the clusters form, rather than the
global environment of the galaxy itself.

We suggest follow-up observations for galaxies as a function of
environment, especially isolated giant elliptical galaxies.  Also, it
is statistically difficult to see any MMRs in spiral galaxies, but if
they could be stacked together in a similar manner to the dwarf
galaxies in \cite{2006ApJ...653..193M} then perhaps a trend would
become apparent.


\acknowledgments
RC and WEH thank the Natural Sciences and Engineering Research Council
of Canada for financial support.  This research has made use of the NASA/IPAC Extragalactic Database (NED) which is operated by the Jet Propulsion Laboratory, California Institute of Technology, under contract with the National Aeronautics and Space Administration.

\section{Appendix}
\label{appendix}

To estimate the total number of GCs in the Gemini systems, we first counted the
number of GCs within similar limits of Section \ref{bimodality}:
\begin{itemize}
\item 0.5 $<$ $(g'-i')_0$ $<$ 1.4 and M$_{i'}$ $<$ -8.35 for NGC 5193
  and 
\item 0.25 $<$ $(g'-i')_0$ $<$ 1.25 and M$_{i'}$ $<$ -8.4 for IC 4329.
\end{itemize}

These magnitudes were converted to M$_V$ with the \textit{ugriz}
standard star catalogue by E. Wehner referenced in Section
\ref{calibration}.  \textit{(i'-V)} was plotted against
\textit{(g'-i')} 
and the following conversion equation was obtained:

\begin{equation}
(i'-V)=-0.69(g'-i')+0.8238
\end{equation}

\noindent 

\begin{figure}[!]
  \begin{center}
    \includegraphics[width=75mm]{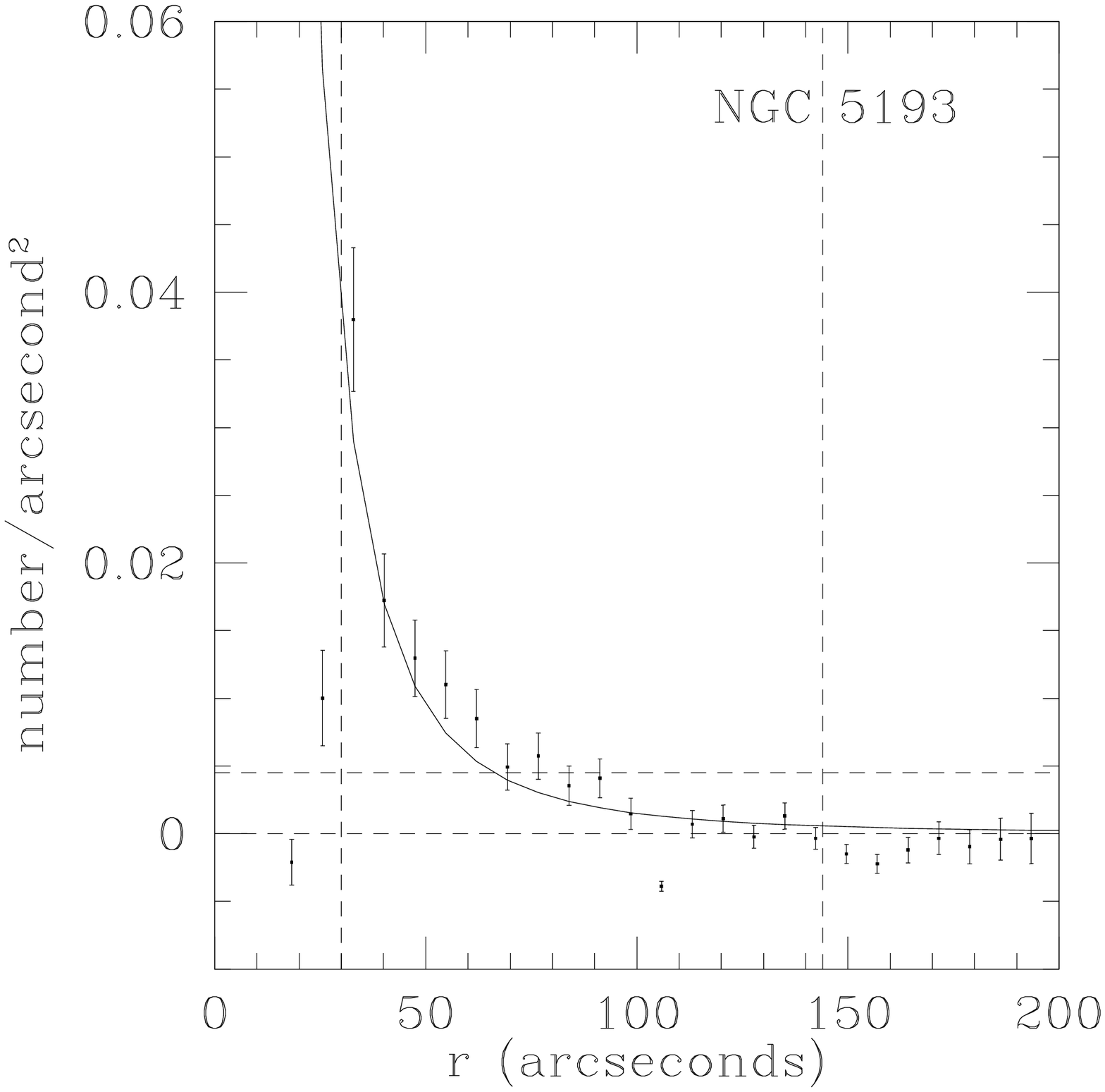}
    \includegraphics[width=75mm]{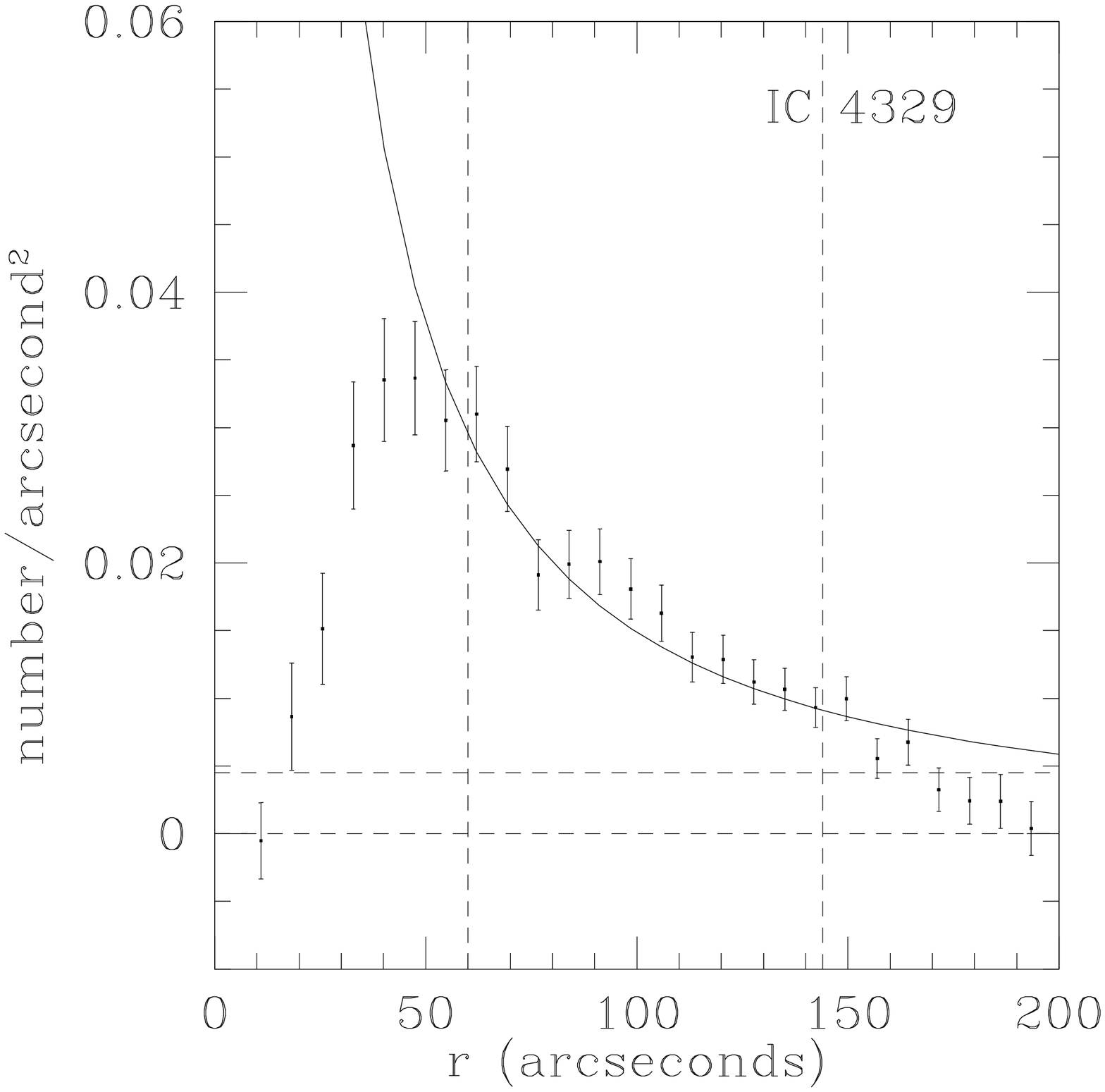}
    \caption{Number of GC candidates per unit area on the sky as a function
of projected galactocentric radius, after background subtraction.
The radial bins used were 50 pixels wide, corresponding to
7.3 arcseconds.  NGC 5193 and IC 4329 have scales of 3.64 and 2.85
arcseconds/kpc, respectively.  The error bars equal ($n^{0.5}$/area) for $n$
candidates in each bin.  The upper horizontal dashed line in
each panel shows the adopted background level $\sigma_b$, chosen
from the outer regions of the NGC 5193 field where the observed
number density is nearly level.  The power-law fits described
in the Appendix were determined from the data points within
the two vertical dashed lines: the inner one represents the
point within which the detections of the GC candidates become
hampered by the increasing light of the central bulge, whereas
the outer line represents the largest radius at which a complete
annulus fits inside the GMOS field.}
    \label{radialdist}
  \end{center}
\end{figure}

\begin{figure}[h]
  \begin{center}
  \includegraphics[width=90mm]{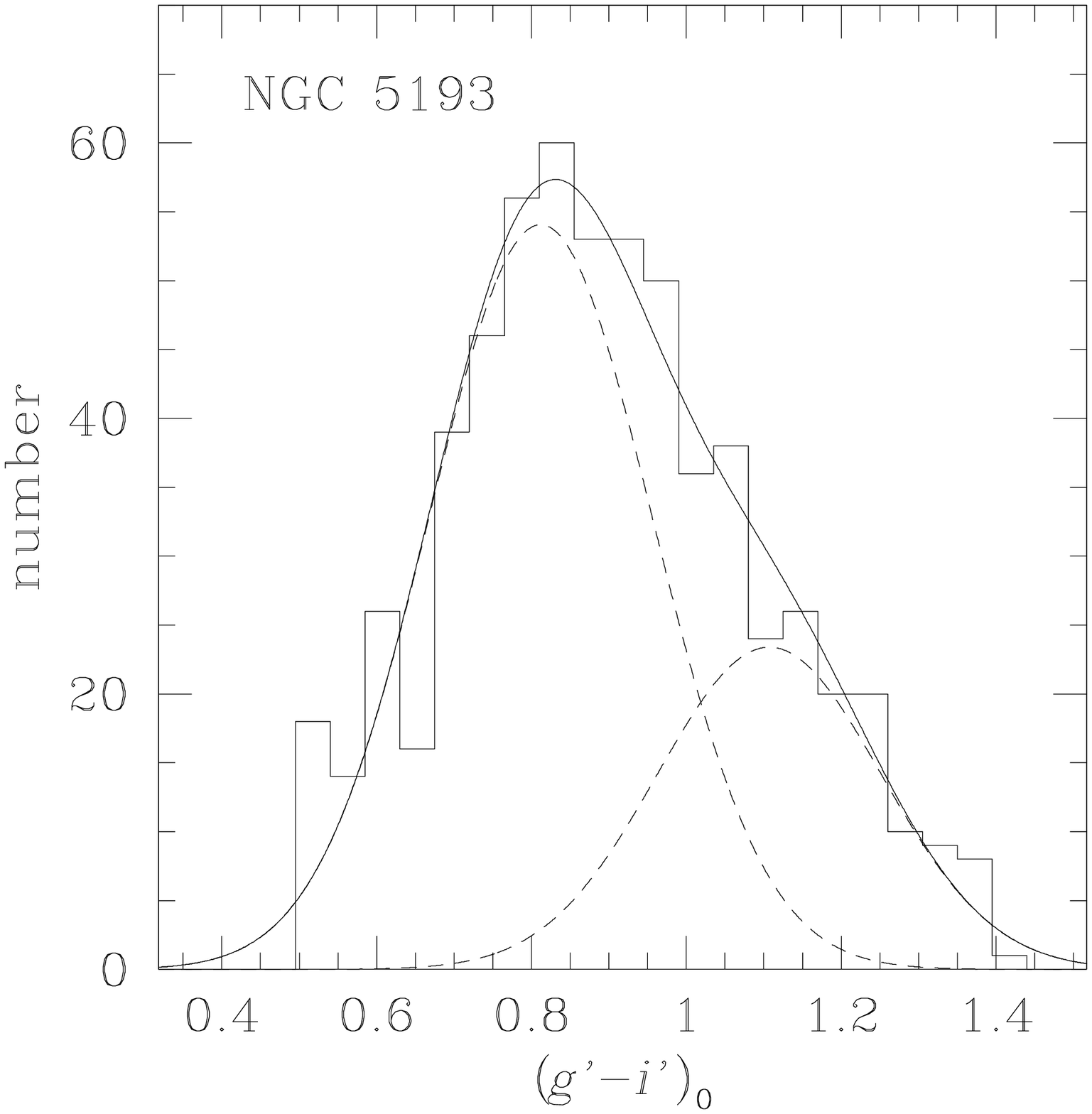}
  \includegraphics[width=90mm]{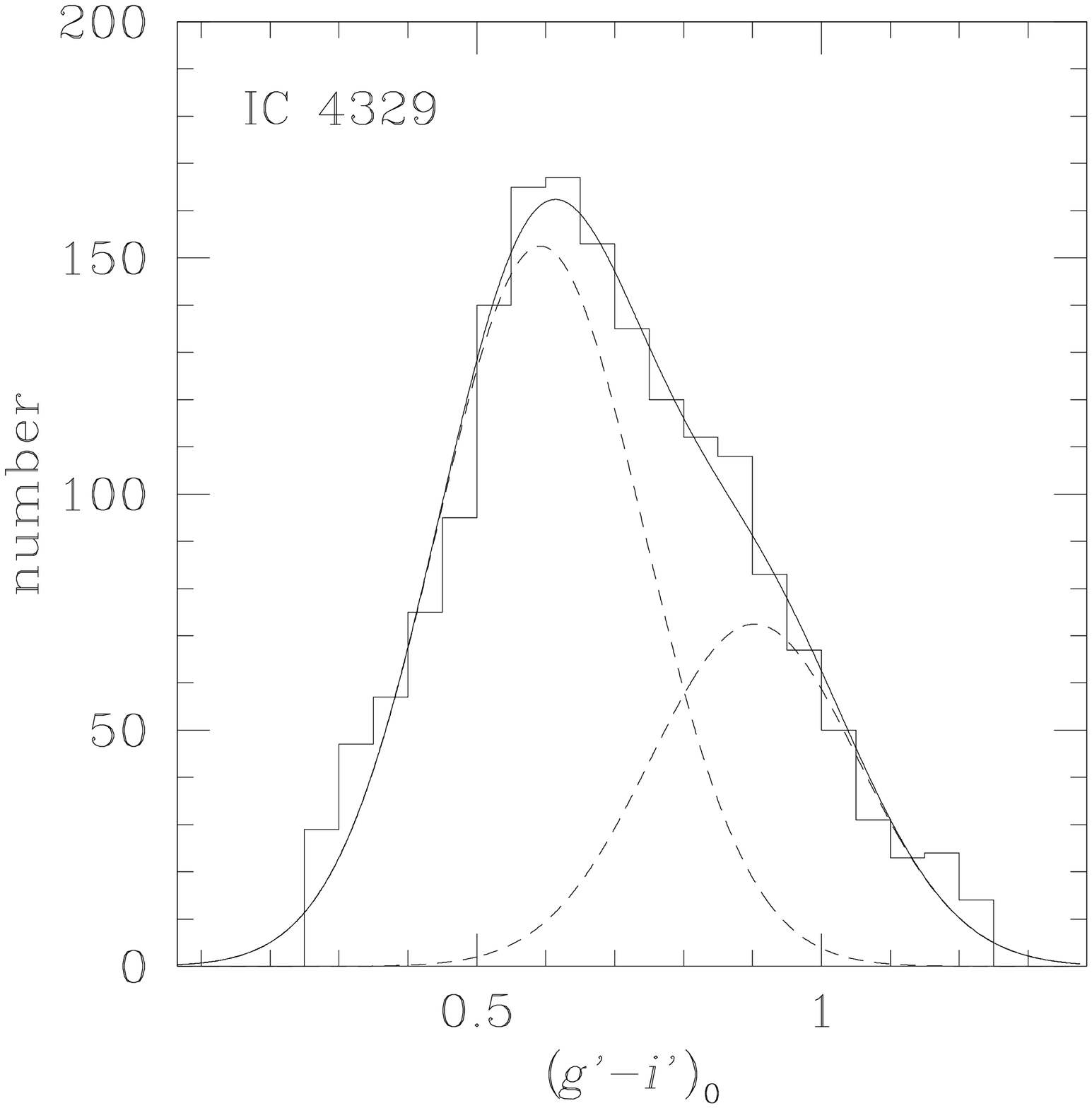}
  \end{center}
  \caption{The two-Gaussian solutions for the RMIX fits of NGC 5193
    and IC 4329 data sets.  The parameters of these fits are shown in
    Table \ref{rmixparams}.  We note that although the distributions
    look like they could be fit with a skewed unimodal solution, we
    assume bimodality from previous studies (e.g.,
\citealt{2006ApJ...636...90H, 2006ApJ...653..193M,
  2006AJ....132.1593S, 2006AJ....132.2333S}).} 
  \label{rmix}
\end{figure}

\begin{figure}[!]
  \includegraphics[width=150mm]{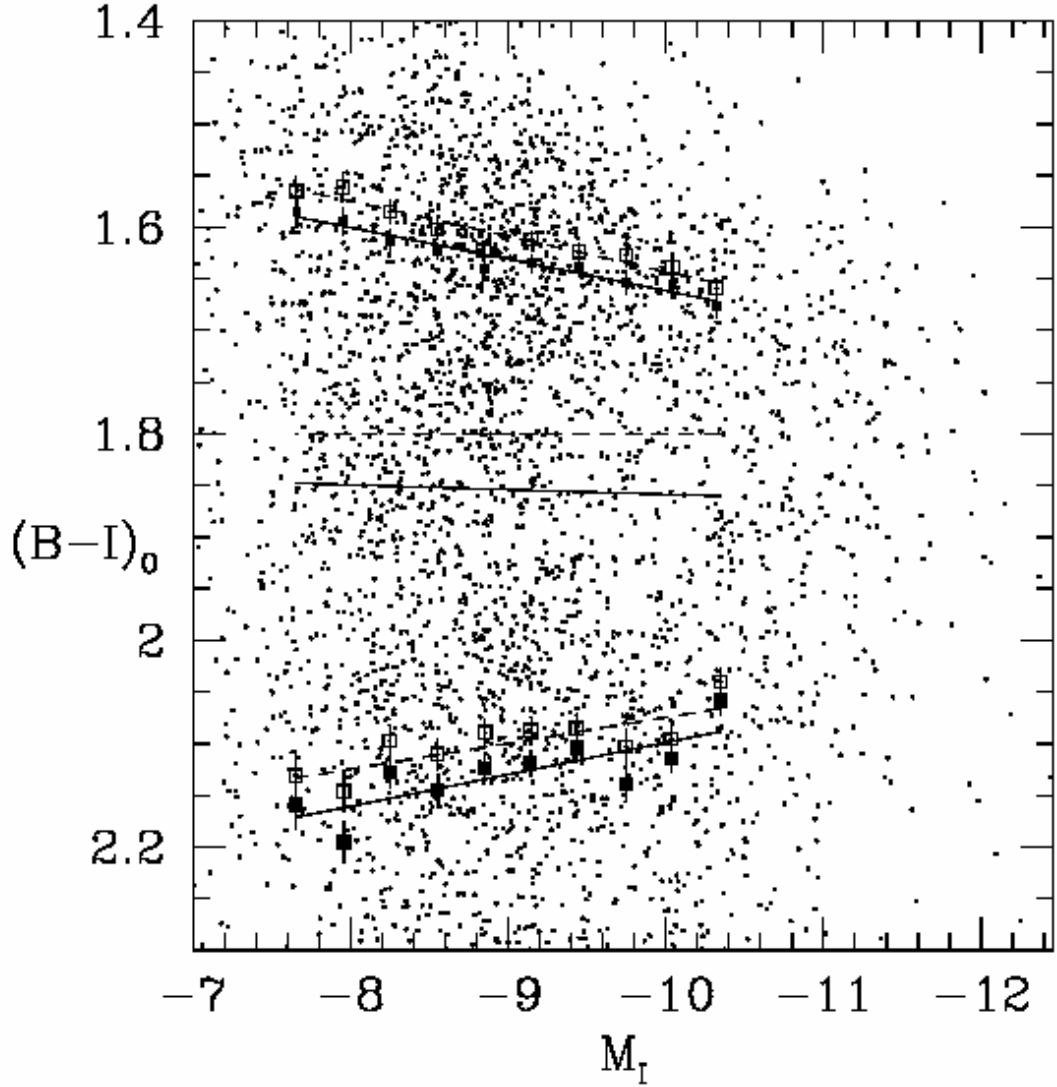}
  \caption[A CMD for NGC 4696, with an example of initial constant
  dividing line and a subsequent sloping dividing
  line]{A CMD for NGC 4696, with an example of initial constant dividing line with a value of
    \textit{(B-I)$_0$}=1.8 (the middle dashed line) and then a
    subsequent sloping dividing
  line (the middle solid line). The sloping dividing line is placed
  exactly half-way between the initial WLSQ fits (the upper and lower
  dashed lines).  See Section \ref{mmr}.}
  \label{initialsplit}
\end{figure}

\begin{figure}
  \begin{center}
    \includegraphics[width=90mm]{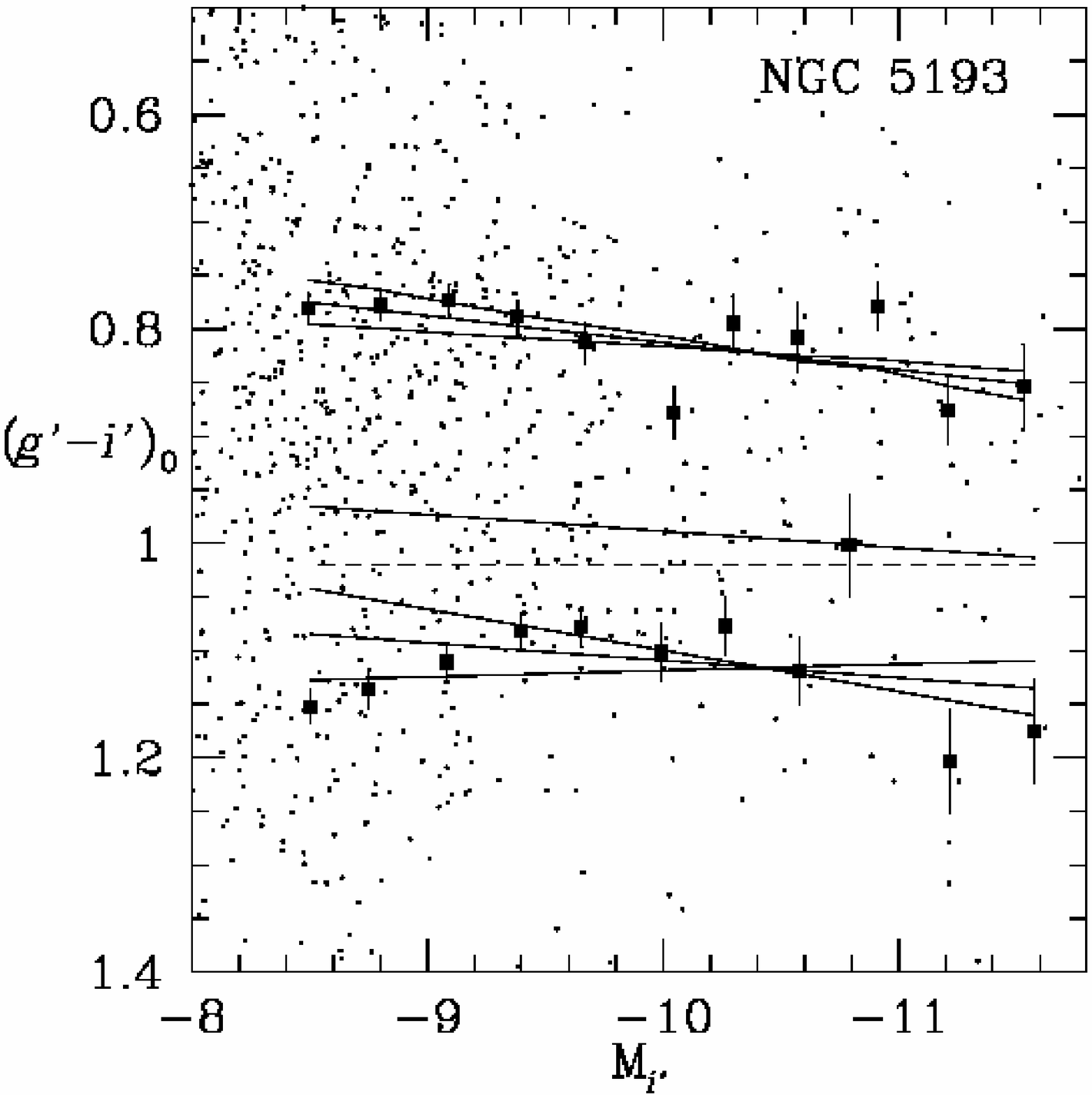}
    \includegraphics[width=90mm]{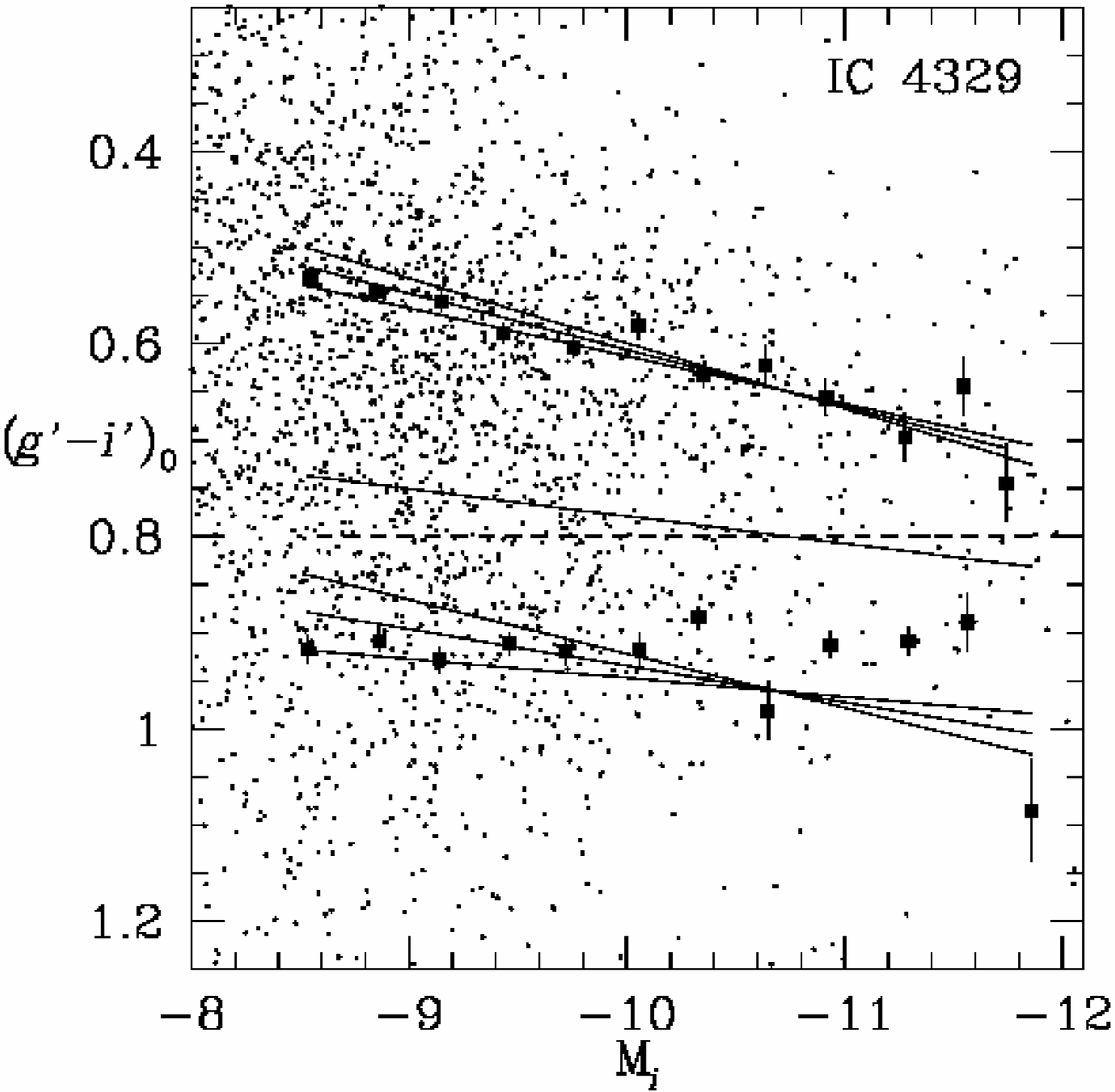}
  \end{center}
  \caption{WLSQ fits and their one-sigma errors for the blue and red
    subpopulations for NGC 5193 and IC 4329.  The dashed line shows
    the initial splitting line and the solid line represents the
    second splitting line.  This solid splitting line results in the blue (red) WLSQ
    fit and error lines above (below) it.}
  \label{51934329mmrs}
\end{figure}

\begin{figure}[!]
  \begin{center}
    \includegraphics[width=80mm]{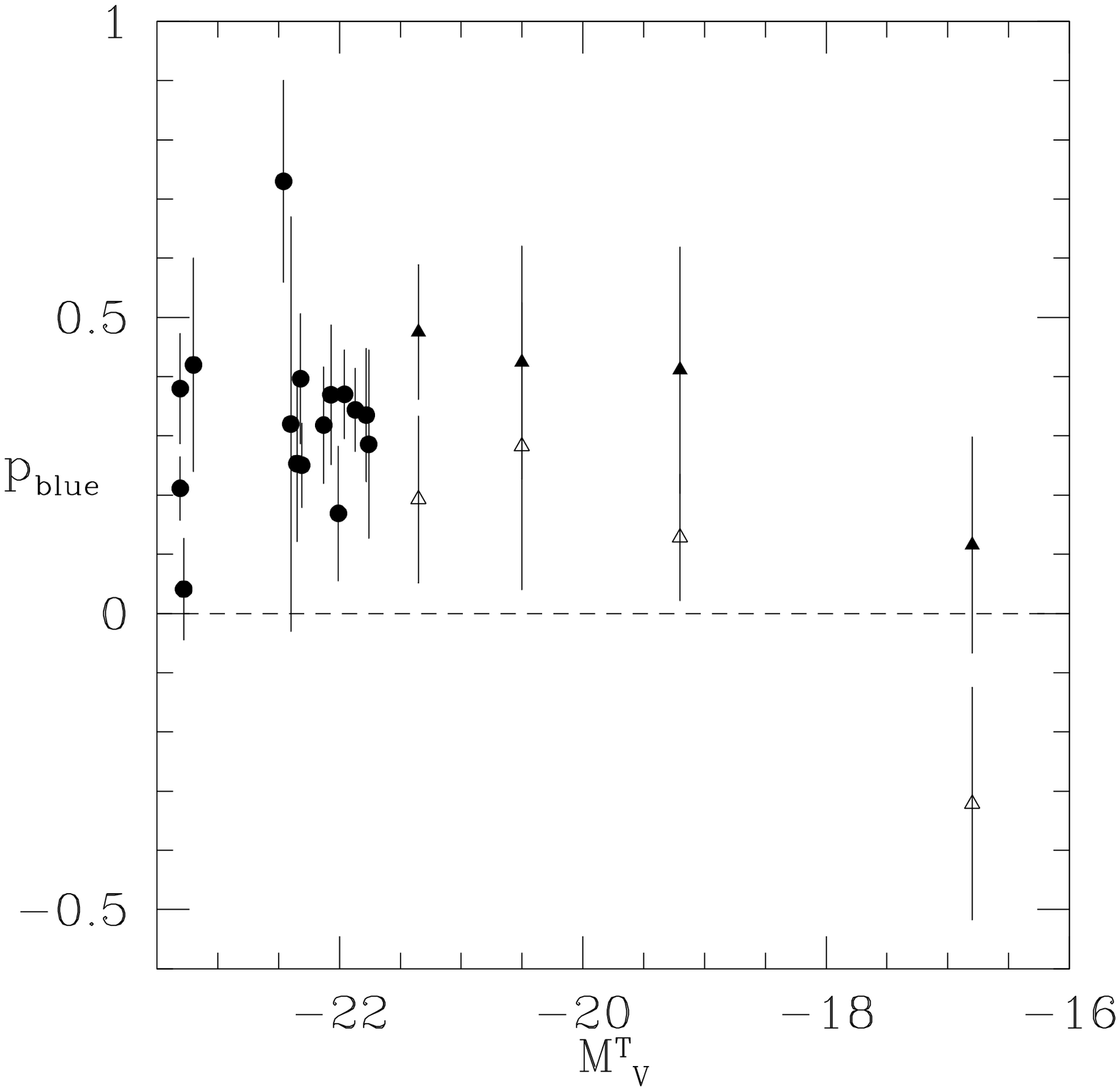}
    \includegraphics[width=80mm]{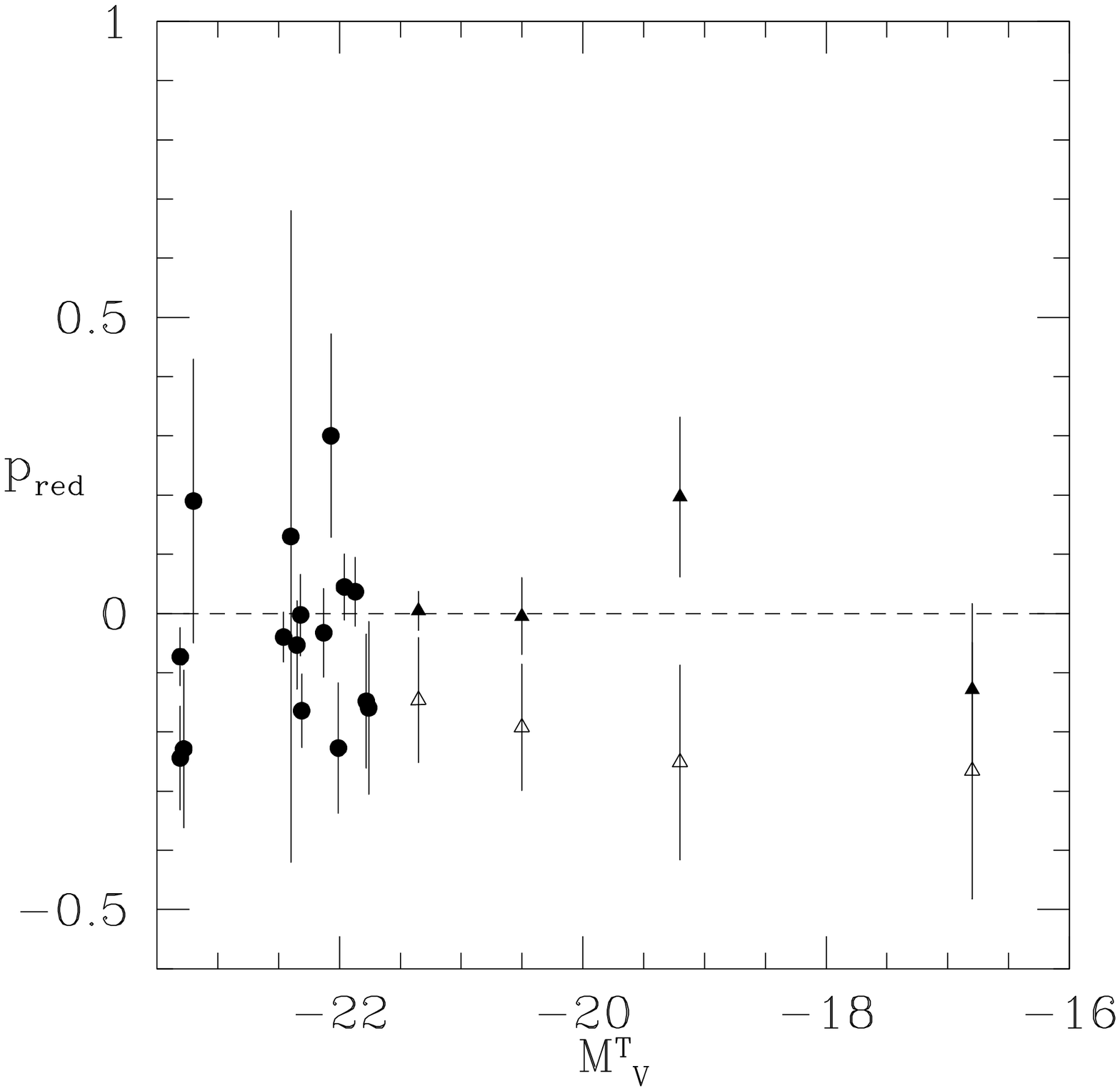}
    \caption[Mean of blue and red slopes of metallicity, zoomed
    out]{Mean of blue and red MMR indices plotted
      against host-galaxy luminosity.  For our data sets (filled circles), the one-sigma errors that were calculated for the colour-magnitude
      slopes are shown as the same relative error on $p$ here.
      Transformation errors are not included here but are
      approximately $\Delta p \approx$0.02 and 0.2 for 
      $(B-I)_0$ to metallicity transformations and $(g'-i')$ to
      metallicity transformations, respectively.  The data points of
      \cite{2006ApJ...653..193M} are shown here for comparison;
      ($g-z,z$) and ($g-z,g$) as filled and open triangles,
      respectively.  The errors for their points come from a random
      resampling around the KMM fits (see their Table 1).  Note that although we tested 15 GCSs, there are actually
      16 points plotted above.  The two data points at $M^T_V$=-23 are both for NGC 4696, with the upper
      (lower) data point corresponding to the method using PSF
      (aperture) photometry \citep{2006ApJ...636...90H}.}
    \label{comparisonzoomedout}
  \end{center}
\end{figure}

\begin{figure}
  \begin{center}
    \includegraphics[width=90mm]{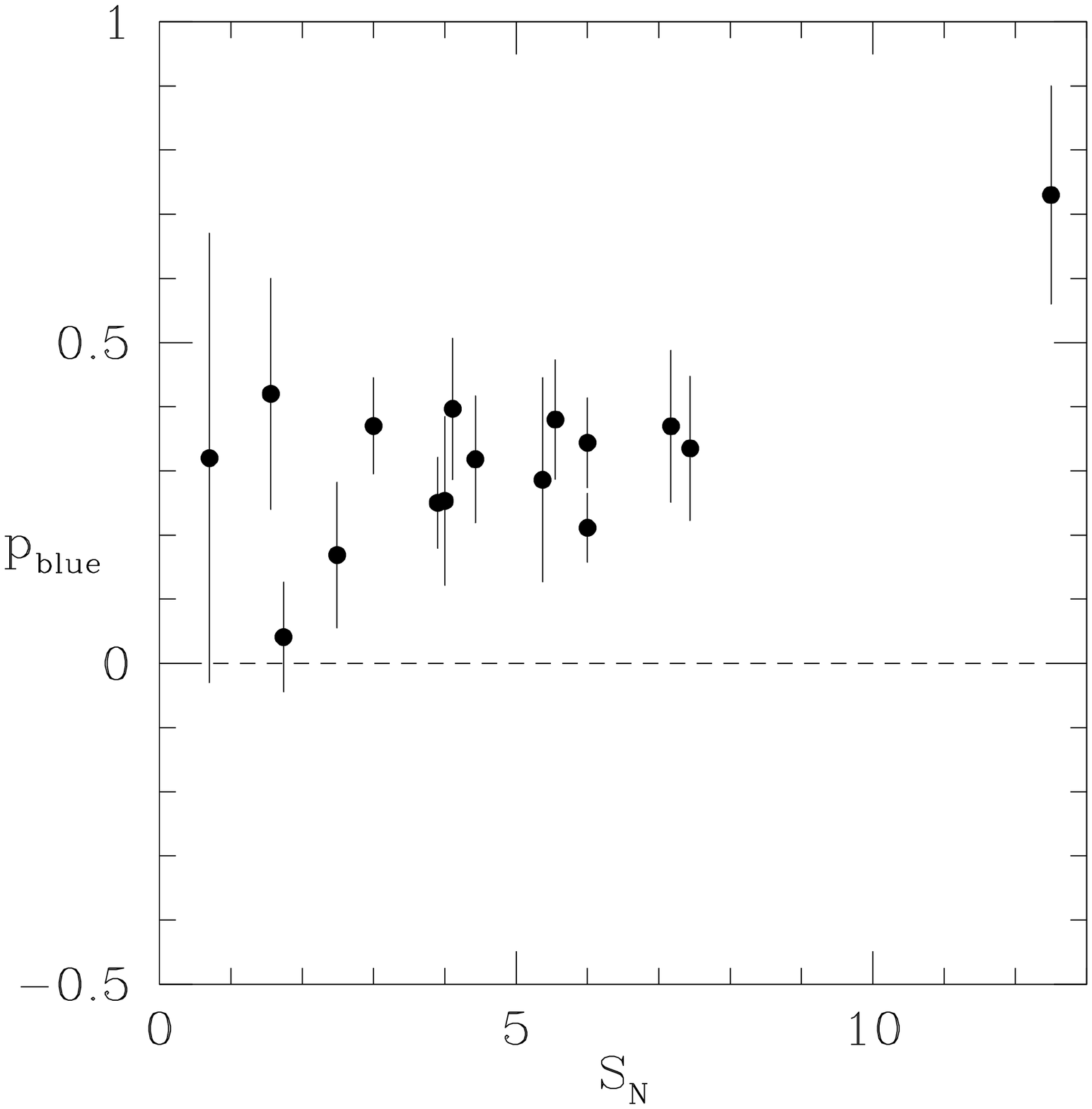}
    \includegraphics[width=90mm]{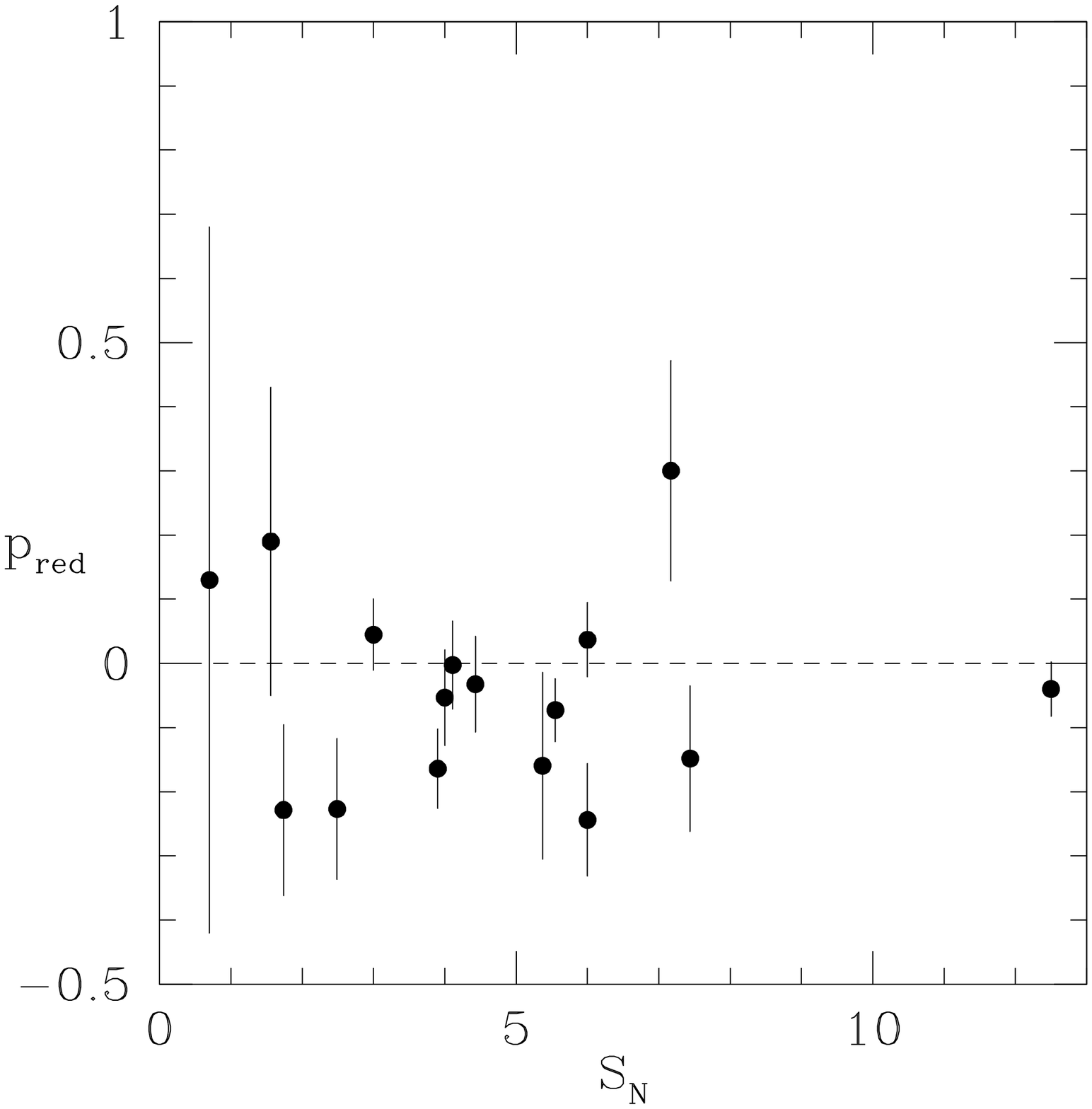}
  \end{center}
  \caption{Mean of blue and red MMR indices plotted
    against the specific frequency, $S_N$.  Values and errors for the
    indices are the same as those in Figure \ref{comparisonzoomedout}.
   As with Figure \ref{comparisonzoomedout} we note that although we
   tested 15 GCSs, there are actually 16 points plotted above due to
   both PSF and aperture photometry results being included for NGC
   4696 \citep{2006ApJ...636...90H}.}
  \label{slopessn}
\end{figure}

\begin{figure}
  \begin{center}
    \includegraphics[width=95mm]{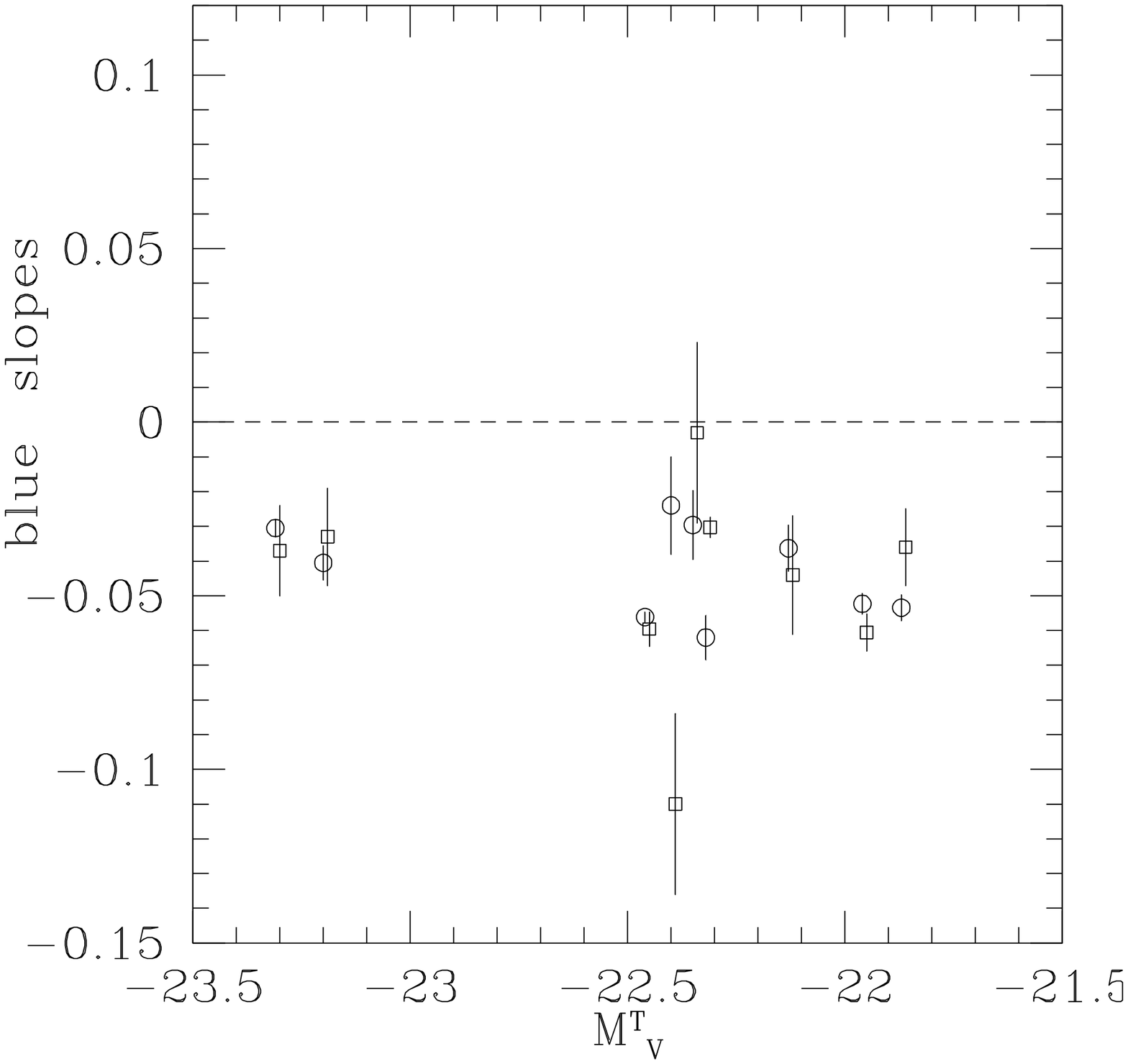}
    \includegraphics[width=95mm]{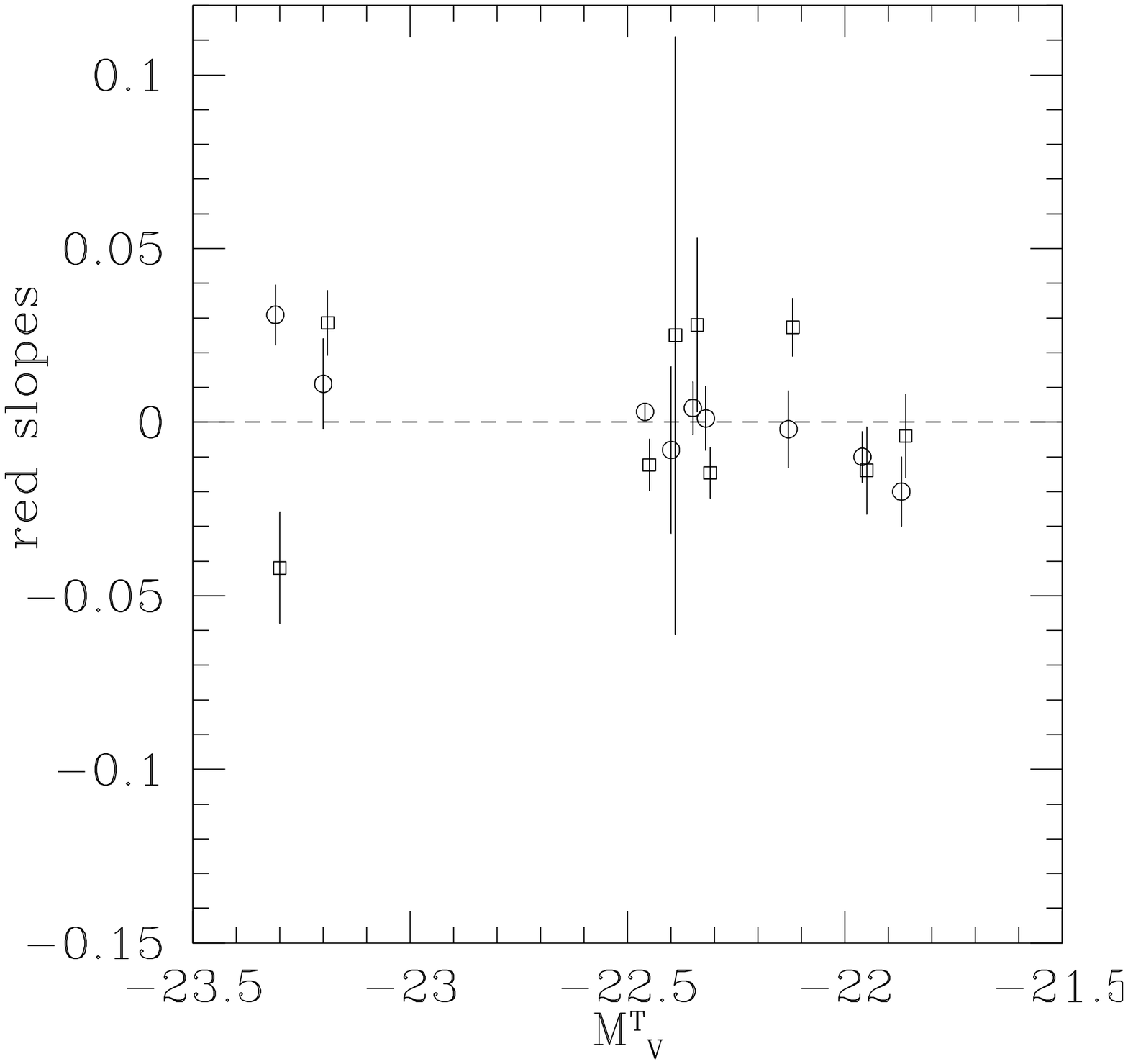}
  \end{center}
  \caption[Comparing line split slopes with slopes from peaks of KMM
    and RMIX fits]{Comparing line split colour slopes with colour slopes from peaks of KMM
    and RMIX fits as in Table \ref{slopevskmmtable}.  Each GCS has two
    data points associated with it; line split
    slopes are on the left (open circles), with KMM/RMIX slopes offset from the true
    host-galaxy magnitude to the right (open squares).}
  \label{comparingbrslopes}
\end{figure}

The radial distribution of the objects falling in these magnitude
and colour ranges, relative to the centre of each galaxy, is shown
in Figure \ref{radialdist}.  Here the number density (objects per unit area)
is plotted against galactocentric distance in linear form,
extending out to the edges of the GMOS frame.  The outer regions
of the NGC 5193 field were used to set an approximate background
level for both fields and this was subtracted from the counts
before plotting.  Then, to estimate the total GC population within
the field, we fit a simple power law of the form  $\sigma = \beta
r^{-\alpha}$ and integrated out to the edge of the image.  The
power laws, after being amplitude adjusted, were $\sigma = 313.2\pm
50.4 r^{-2.66\pm0.22}$ and $\sigma = 7.12\pm1.52 r^{-1.34\pm0.10}$ for
NGC 5193 and IC 4329, respectively.  As this estimate does not
cover the entire radial extent of the GCS, we corrected for the field
of view in our images.  

We then assumed that for a gE the number of GCs per unit
magnitude is described by a Gaussian with a mean value of M$_V$ =
-7.33 and a standard deviation of 1.4 magnitudes
\citep{2001stcl.conf..223H}.  This allowed us to find the fraction of GCs brighter than the limiting
magnitude.  By dividing the number of GCs above the limiting magnitude
by this fraction, the estimate for the total number of GCs in
the system was finally obtained (i.e., the values in column (8) in Table
\ref{galaxydetails}).  The specific frequency, S$_N$ is then calculated
from 

\begin{equation}
S_N=N_{GC}10^{0.4(M_V+15)}
\end{equation}

\noindent \citep{1981AJ.....86.1627H}.  For the remaining galaxies, we
either adopted values of $S_N$ from the literature or, if there were
no previous values, we calculated $S_N$
in a method similar to the above without the FOV correction.  For $HST$ data, limiting magnitudes were converted to
M$_V$ with \textit{UBVRI} standard star data from
\cite{1992AJ....104..340L}.  \textit{(V-I)} was plotted against
\textit{(B-I)} 
and the following conversion equation was obtained:

\begin{equation}
(V-I)=0.53(B-I).
\end{equation}

\bibliography{adsbibliography.bib}


\begin{deluxetable}{cccc}
  \tablecaption{Summary of MMRs in Previously Observed Galaxies
    \label{previousmmrs}}
  \tabletypesize{\scriptsize}
  \tablewidth{0pt}     
  \tablehead
  {
    \colhead{Galaxy sample}&\colhead{$p$}&\colhead{Brightest GC
      Luminosity$^{\rm a}$}&\colhead{Reference}\\
    (1)&(2)&(3)&(4)\\
  }
  \startdata
  8 BCGs&0.55&M$_I$$>$-13.0&\cite{2006ApJ...636...90H}\\
  NGC 3311&0.6&M$_I$$>$-13.5&\cite{2008ApJ...681.1233W}\\
  Virgo galaxies{$^{\rm b}$}&0.48$\pm$0.08&M$_Z$$>$-12.0&\cite{2006ApJ...653..193M}\\
  M87&0.48&M$_Z$$>$-12.2&\cite{2006AJ....132.2333S}\\
  NGC 4649&0.43&M$_Z$$>$-12.0&\cite{2006AJ....132.2333S}\\
  NGC 4472 (M49)&0&M$_Z$$>$-11.8&\cite{2006AJ....132.2333S}\\
  NGC 5866&0.11$\pm$0.09{$^{\rm c}$}&M$_R$$>$-10.2&\cite{2007ApJ...668..209C}\\
  NGC 1533&0&M$_{I(814)}$$>$-10.2&\cite{2007ApJ...671.1624D}\\
  NGC 4594 (Sombrero Galaxy)&0.27&M$_R$$>$-11.6&\cite{2006AJ....132.1593S}\\
\hline
  \multicolumn{4}{l}{$^{\rm a}$ Values of \textit{cz} and extinction
    from NED were used with an adopted value of}\\
  \multicolumn{4}{l}{\textit{H$_0$} = 70 km s$^{-1}$ Mpc$^{-1}$ to
    calculate absolute magnitudes for M87, NGC 4649, NGC 4472,}\\
  \multicolumn{4}{l}{NGC 5866, NGC 1533 and NGC 4594.}\\
  \multicolumn{4}{l}{$^{\rm b}$ \cite{2006ApJ...653..193M} study 79
    early type galaxies which they split into four host-galaxy}\\
  \multicolumn{4}{l}{magnitude bins.  The MMR cited above is for their brightest bin where the
    host-galaxy magnitude}\\
  \multicolumn{4}{l}{range is -21.7$<$M$_B$$<$-21.}\\
  \multicolumn{4}{l}{$^{\rm c}$ \cite{2007ApJ...668..209C} caution
      that this result is tentative because of the small number of}\\ 
      \multicolumn{4}{l}{GCs (109) used to find these slopes.}\\
  \enddata
\end{deluxetable}

\begin{deluxetable}{ccccccccc}
  \tablecaption{Data for Individual BCG Galaxies
    \label{galaxydetails}}
  \tabletypesize{\scriptsize}
  \tablewidth{0pt}     
  \tablehead
  {
    \colhead{NGC or IC}&\colhead{Cluster or Group}&\colhead{Redshift \textit{cz}}&\colhead{\textit{M{$_V^T$}}}&\colhead{\textit{E(B-I)}}&\colhead{\textit{(m-M){$_I$}}}&\colhead{Number of}&\colhead{N$_{GC}$}&\colhead{S$_{N}$}\\
    & &(km s{$^{-1}$})& & & &data points& &\\
    (1)&(2)&(3)&(4)&(5){$^{\rm a}$}&(6){$^{\rm b}$}&(7){$^{\rm c}$}&(8)&(9)\\
   }
  \startdata
  \multicolumn{9}{c}{\textit{(B, I)} photometry from ACS/WFC}\\
  \hline
  NGC 1407&Eridanus&1627{$^{\rm i}$}&-22.4&0.16&32.0&1046&2641$\pm$303{$^{\rm d}$}&4.0$\pm$1.3{$^{\rm d}$}\\
  NGC 5322&CfA 122&1916{$^{\rm i}$}&-22.0&0.03&32.2&395&1600&2.5\\
  NGC 7049&N7049&1977{$^{\rm i}$}&-21.8&0.12&32.4&682&2700&5.8\\
  NGC 3348&CfA 69&2902{$^{\rm a}$}&-22.2&0.17&33.2&831&3100&4.4\\
  NGC 3258&Antlia&3129{$^{\rm i}$}&-22.1&0.20&33.4&1812&6000$\pm$150{$^{\rm e}$}&6$\pm$2.5{$^{\rm e}$}\\
  NGC 3268&Antlia&3084{$^{\rm i}$}&-22.1&0.24&33.4&1593&4750$\pm$150{$^{\rm e}$}&3$\pm$2{$^{\rm e}$}\\
  NGC 4696&Cen 30&3248{$^{\rm i}$}&-23.0&0.23&33.5&3099&4100$\pm$200{$^{\rm f}$}&6{$^{\rm f}$}\\
  NGC 5557&CfA 141&3389{$^{\rm a}$}&-22.4&0.01&33.4&841&3500&4.1\\
  NGC 7626&Pegasus (I)&3154{$^{\rm j}$}&-22.4&0.17&33.4&1835&2833$\pm$300{$^{\rm g}$}&3.9{$^{\rm g}$}\\
  \hline
  \multicolumn{9}{c}{\textit{B} photometry from ACS/WFC, \textit{I} from WFPC2}\\
  \hline
  NGC 708&Abell 262&4601{$^{\rm a}$}&-22.1&0.21&34.3&1282&4800&7.2\\
  NGC 7014&Abell 3742&4676{$^{\rm a}$}&-21.8&0.08&34.2&962&3800&7.4\\
  IC 4296&Abell 3565&4009{$^{\rm a}$}&-23.4&0.15&33.8&885&36000&1.7\\
  \hline
  \multicolumn{9}{c}{\textit{(g, i)} photometry from GMOS}\\
  \hline
  NGC 5193&Abell 3565&3988{$^{\rm a}$}&-22.4&0.11&33.9&596&630$\pm$330&0.70$\pm$0.36\\
  IC 4329&Abell 3574&4808{$^{\rm a}$}&-23.1&0.12&34.3&1528&2970$\pm$530&1.56$\pm$0.28\\
  NGC 3311&Abell 1060&3937{$^{\rm a}$}&-22.4&0.15&34.0&4936&16500$\pm$2000{$^{\rm h}$}&12.5$\pm$1.5{$^{\rm h}$}\\
  \hline

  \multicolumn{9}{l}{$^{\rm a}$ Values were taken from NED.}\\
  \multicolumn{9}{l}{$^{\rm b}$ Values calculated using an adopted
    value of \textit{H}$_0$ = 70 km s$^{-1}$ Mpc$^{-1}$.}\\
  \multicolumn{9}{l}{$^{\rm c}$ Points in the initial selection
    regions as detailed in Section \ref{bimodality}}\\
  \multicolumn{9}{l}{Values from}\\
  \multicolumn{9}{l}{$^{\rm d}$ \cite{1997AJ....113..895P}.}\\
  \multicolumn{9}{l}{$^{\rm e}$ \cite{2008MNRAS.tmp..421B}.}\\
  \multicolumn{9}{l}{$^{\rm f}$ \cite{1999IAUS..186..200L}.}\\
  \multicolumn{9}{l}{$^{\rm g}$ \cite{2006A&A...458...53S}.}\\
  \multicolumn{9}{l}{$^{\rm h}$ \cite{2008ApJ...681.1233W}.}\\
  \multicolumn{9}{l}{$^{\rm i}$ \cite{2001ApJ...546..681T}.}\\
  \multicolumn{9}{l}{$^{\rm j}$ \cite{2002AJ....123.2976R}.}\\
  \enddata
\end{deluxetable}

\begin{deluxetable}{cccccccccc}
  \tablecaption{Blue and red
    $\Delta$\textit{(B-I)$_0$}/$\Delta$\textit{M}$_I$ slopes with corresponding \textit{p} values for all bins, where \textit{p} is defined in the relation
    \textit{Z} = \textit{M$^p$}.  Errors are shown as the internal
    relative error of the original slopes (i.e., errors due to the
    colour transformation are not propagated).  The IC 4329 slopes for
    12 and 13 bins are shown with and without the outlying point
    mentioned in Section \ref{mmr}.
    \label{redblueslopestable}}
  \tabletypesize{\scriptsize}
  \tablewidth{0pt}     
  \tablehead
  {
    \colhead{NGC}&\colhead{Number}&\colhead{Blue}&\colhead{Blue
      Slope}&\colhead{p}&\colhead{Error}&\colhead{Red}&\colhead{Red Slope}&\colhead{p}&\colhead{Error}\\
    or IC&of bins&Slope$^{\rm a}$&Error$^{\rm a}$& &on p&Slope$^{\rm a}$&Error$^{\rm a}$& &on p\\
    (1)&(2)&(3)&(4)&(5)&(6)&(7)&(8)&(9)&(10)\\    
  }
  \startdata
  NGC 5193&10&-0.05&0.03&0.31&0.56&-0.01&0.05&0.05&0.78\\
  NGC 5193&11&-0.05&0.02&0.32&0.45&-0.03&0.04&0.21&0.77\\
  \hline
  IC 4329&10&-0.08&0.01&0.51&0.32&0.02&0.02&-0.14&0.44\\
  IC 4329&11&-0.05&0.01&0.35&0.33&0.05&0.02&-0.33&0.44\\
  IC 4329(-outlier)&12&-0.11(-0.111)&0.02(0.003)&0.74(0.74)&0.53(0.30)&-0.07(0.01)&0.03(0.02)&0.48(-0.04)&0.73(0.36)\\
  IC 4329(-outlier)&13&-0.08(-0.082)&0.01(0.003)&0.55(0.55)&0.42(0.23)&-0.06(0.01)&0.03(0.02)&0.42(-0.06)&0.64(0.38)\\
  \hline
  NGC 3311&10&-0.106&0.003&0.71&0.28&0.006&0.005&-0.038&0.090\\
  NGC 3311&11&-0.103&0.003&0.69&0.27&0.007&0.004&-0.050&0.080\\
  NGC 3311&12&-0.118&0.006&0.79&0.36&0.005&0.003&-0.033&0.059\\
  \hline
  \multicolumn{10}{l}{$^{\rm a}$ Values converted from \textit{(g-i)$_0$} to \textit{(B-I)$_0$}}\\
  \enddata
\end{deluxetable}

\begin{deluxetable}{ccccccccccc}
  \tablecaption{RMIX Multimodal Fitting Parameters for Gemini/GMOS Galaxies
    \label{rmixparams}}
  \tabletypesize{\scriptsize}
  \tablewidth{0pt}     
  \tablehead
  {
    \colhead{Target}&\colhead{Pi}&\colhead{Error}&\colhead{Mean}&\colhead{Error}&\colhead{Sigma}&\colhead{Error}&\colhead{DOF}&\colhead{$\chi^2$}&\colhead{Reduced}&\colhead{pr($>\chi^2$)}\\
    & & & & & & & & &$\chi^2$& \\
    (1)&(2)&(3)&(4)&(5)&(6)&(7)&(8)&(9)&(10)&(11)\\    
  }
  \startdata
      NGC 5193&0.70&0.05&0.81&0.01&0.144&0.007&19&46.1&2.4&2.2x10$^{-16}$\\
       &0.30&0.05&1.11&0.02&0.144&NA& & & & \\
      \hline
      IC 4329&0.68&0.03&0.592&0.009&0.150&0.005&18&69.8&3.9&4.8$^{-4}$\\
       &0.32&0.03&0.90&0.01&0.150&NA& & & & \\
      \hline
      NGC 3311&0.57&0.02&0.610&0.005&0.142&0.003&19&221&11.6&4.8x10$^{-8}$\\
       &0.43&0.02&0.898&0.006&0.142&NA& & & & \\
  \enddata
\end{deluxetable}

\begin{deluxetable}{ccccccccccc}
  \tablecaption{Comparison of Line Split $\Delta(B-I)_0/\Delta M_I$
    Slopes For Blue and Red GC Sequences.  Pairs that do not agree
    within the errors are highlighted in bold.  The slopes for the
    line split bins are from the WLSQ fits that have the maximum
    number of bins minus the faintest and brightest bins.
    \label{slopevskmmtable}}
  \tabletypesize{\scriptsize}
  \tablewidth{0pt}     
  \tablehead
  {
    \colhead{Galaxy}&\colhead{Method}&\colhead{Blue}&\colhead{Error}&\colhead{Red}&\colhead{Error}\\
    & &Slope& &Slope& \\
    (1)&(2)&(3)&(4)&(5)&(6)\\    
  }
  \startdata
  NGC 1407&Line split bins&{\bf -0.041}&{\bf 0.011}&0.0168&0.0077\\
  &KMM peaks&{\bf 0.00}&{\bf 0.03}&0.03&0.03\\
  \hline
  NGC 3348&Line split bins&-0.0517&0.0081&0.0148&0.0065\\
  &KMM peaks&-0.04&0.02&0.027&0.008\\
  \hline
  NGC 3258&Line split bins&-0.0495&0.0027&0.0083&0.0064\\
  &KMM peaks&-0.04&0.01&0.00&0.01\\
  \hline
  NGC 3268&Line split bins&-0.0571&0.0030&{\bf 0.0113}&{\bf 0.0075}\\
  &KMM peaks&-0.061&0.005&{\bf -0.01}&{\bf 0.01}\\
  \hline
  NGC 4696&Line split bins&-0.0314&0.0040&{\bf 0.0486}&{\bf 0.070}\\
  (aperture)&KMM peaks&-0.04&0.01&{\bf -0.04}&{\bf 0.02}\\
  \hline
  NGC 5557&Line split bins&{\bf -0.0498}&{\bf 0.0054}&-0.0035&0.0077\\
  &KMM peaks&{\bf -0.030}&{\bf 0.003}&-0.015&0.007\\
  \hline
  NGC 5193$^{\rm a}$&Line split bins&{\bf -0.025}&{\bf 0.018}&-0.012&0.032\\
  &RMIX peaks&{\bf -0.11}&{\bf 0.03}&0.03&0.09\\
  \hline
  IC 4329$^{\rm a}$&Line split bins&-0.0400&0.0089&-0.033&0.021\\
  &RMIX peaks&-0.03&0.01&0.03&0.01\\
  \hline
  NGC 3311$^{\rm a}$&Line split bins&-0.0545&0.0012&{\bf 0.0038}&{\bf 0.0023}\\
  &RMIX peaks&-0.060&0.005&{\bf -0.012}&{\bf 0.007}\\
  \hline 
  \multicolumn{6}{l}{$^{\rm a}$ Values remain in $(g'-i')_0/M_{i'}$}\\
  \hline
  \enddata
\end{deluxetable}

\clearpage


\end{document}